\documentclass[fleqn,usenatbib]{mnras}

\usepackage{newtxtext,newtxmath}
\usepackage[T1]{fontenc}
\usepackage{ae,aecompl}

\usepackage{graphicx}	% Including figure files
\usepackage{amsmath}	% Advanced maths commands
\usepackage{amssymb}	% Extra maths symbols

\title[Detection of HC$_3$NH$^+$ and HCNH$^+$ in L1544]{Detection of the HC$_3$NH$^+$ and HCNH$^+$ ions in the L1544 pre-stellar core}

\author[D. Qu\'enard et al.]
{
D. Qu\'enard,$^{1,2,3}$\thanks{E-mail: d.quenard@qmul.ac.uk}
C. Vastel,$^{2,3}$
C. Ceccarelli,$^{4}$
P. Hily-Blant,$^{4}$
B. Lefloch,$^{4}$
\newauthor and R. Bachiller$^{5}$
\\
$^{1}$School of Physics and Astronomy, Queen Mary University of London, Mile End Road, London E1 4NS, UK\\
$^{2}$Universit\'e de Toulouse, UPS-OMP, IRAP, Toulouse, France\\
$^{3}$CNRS, IRAP, 9 Av. Colonel Roche, BP 44346, F-31028 Toulouse Cedex 4, France\\
$^{4}$Univ. Grenoble Alpes, CNRS, IPAG, F-38000 Grenoble, France\\
$^{5}$Observatorio Astron\'omico Nacional (OAN, IGN), Calle Alfonso XII, 3, 28014 Madrid, Spain
}

\date{Accepted XXX. Received YYY; in original form ZZZ}

\pubyear{2017}

\begin{document}
\label{firstpage}
\pagerange{\pageref{firstpage}--\pageref{lastpage}}
\maketitle

% Abstract of the paper
\begin{abstract}
{ The L1544 pre-stellar core was observed as part of the ASAI (Astrochemical Surveys At IRAM) Large Program. We report the first detection in a pre-stellar core of the HCNH$^+$ and HC$_3$NH$^+$ ions. The high spectral resolution of the observations allows to resolve the hyperfine structure of HCNH$^+$. Local thermodynamic equilibrium analysis leads to derive a column density equal to (2.0$\pm$0.2)$\times$10$^{13}$\,cm$^{-2}$ for HCNH$^+$ and (1.5$\pm$0.5)$\times$10$^{11}$\,cm$^{-2}$ for HC$_3$NH$^+$. We also present non-LTE analysis of five transitions of HC$_3$N, three transitions of H$^{13}$CN and one transition of HN$^{13}$C, all of them linked to the chemistry of HCNH$^+$ and HC$_3$NH$^+$. We computed for HC$_3$N, HCN, and HNC a column density of (2.0$\pm$0.4)$\times$10$^{13}$\,cm$^{-2}$, (3.6$\pm$0.9)$\times10^{14}$\,cm$^{-2}$, and (3.0$\pm$1.0)$\times$10$^{14}$\,cm$^{-2}$, respectively.
We used the gas-grain chemical code \textit{Nautilus} to predict the abundances all these species across the pre-stellar core. Comparison of the observations with the model predictions suggests that the emission from HCNH$^+$ and HC$_3$NH$^+$ originates in the external layer where non-thermal desorption of other species was previously observed. The observed abundance of both ionic species ([HCNH$^+$]$\,\simeq3\times10^{-10}$ and [HC$_3$NH$^+$]$\,\simeq[1.5-3.0]\times10^{-12}$, with respect to H$_2$) cannot be reproduced at the same time by the chemical modelling, within the error bars of the observations only. We discuss the possible reasons for the discrepancy and suggest that the current chemical models are not fully accurate or complete. However, the modelled abundances are within a factor of three consistent with the observations, considering a late stage of the evolution of the pre-stellar core, compatible with previous observations.}
\end{abstract}

% Select between one and six entries from the list of approved keywords.
% Don't make up new ones.

%%%%%%%%%%%%%%%%%%%%%%%%%%%%%%%%%%%%%%%%%%%%%%%%%%

\section{Introduction}

Protonated hydrogen cyanide, HCNH$^+$, was first discovered in the interstellar medium towards Sgr B2 by \citet{ziurys1986} using the J = 1--0, 2--1, and 3--2 pure rotational transitions at 74, 148, and 222 GHz, respectively. They derived a fractional abundance of 3$\times$10$^{-10}$. Since the initial detection, HCNH$^+$ has also been observed in the TMC-1 dark cloud \citep{schilke1991,ziurys1992} as well as the DR 21(OH) compact HII region \citep{schilke1991, hezareh2008} with abundances of 1.9$\times$10$^{-9}$ and $\sim$ 10$^{-10}$ respectively. This ion appears to be one of the key species in the ion-molecule network of interstellar chemistry. It was also thought to be the main precursor to both HCN and HNC, both formed through rapid dissociative recombination of HCNH$^+$ \citep{herbst1978}.

Protonated Cyanoacetylene (HC$_3$NH$^+$) has been searched towards a sample of dark cloud, evolved star, high-mass star-forming region by \citet{turner1990} with no conclusive detection due to the high rms of their observations. Also, \citet{amano1990} performed a deep search of the 5--4 transition at 43.3\,GHz towards TMC-1 resulting with an upper limit only. HC$_3$NH$^+$ was later detected in the same cloud by \citet{kawaguchi1994} through the 5--4 (43.3 \,GHz) and 4--3 (34.6\,GHz) transitions consistent with an abundance of 1$\times$10$^{-10}$. It is the largest ion detected so far in space. It is thought to play a role in the production of the cyanopolyynes (HC$_{2n+1}$N, where $n$ is an integer, \citealp{turner1990}), which are ubiquitous in cold dark clouds.\\

L1544 is a prototypical pre-stellar core in the Taurus molecular cloud complex (d$\sim$140\,pc) on the verge of gravitational collapse \citep[][and references within]{caselli2002-1, caselli2012}. It is characterised { by a high central density (larger than a few 10$^6$ cm$^{-3}$), a low temperature ($\sim$7\,K), a high CO depletion and a large degree of molecular deuteration \citep{caselli2003, crapsi2005, vastel2006}}. 
Its physical and dynamical structure has been reconstructed by \citet{caselli2012} and \citet{keto2014} { taking into account the numerous observations that have been made} towards L1544 as well as the first detection of water in a pre-stellar core. The observed water line shows an inverse P-Cygni profile \citep{caselli2012, quenard2016}, characteristic of gravitational contraction, confirming that L1544 is on the verge of the collapse. \citet{keto2010} and \citet{keto2014} have shown that L1544 is in slow contraction from unstable quasi-static hydrodynamic equilibrium (an unstable Bonnor-Ebert sphere).\\

As part of the IRAM-30m Large Program ASAI{\footnote{Astrochemical Surveys At Iram: \url{http://www.oan.es/asai/}} ({ Lefloch et al. 2017 in prep.}), { a highly sensitive, unbiased spectral survey of the molecular emission of the L1544 pre-stellar core has been performed with a high spectral resolution. With this survey, \citet{vastel2014} reported the detection of many oxygen bearing complex organic molecules (COMs) and suggested that they were produced through the release in the gas phase of methanol and ethene through non-thermal desorption processes. The sensitivity of these IRAM observations also led to the detection of the hyperfine structure of CH$_2$CN \citep{vastel2015-1} as well as HOCO$^+$ \citep{vastel2016}, likely emitted in the external layer through the same non-thermal desorption processes as for the COMs. In the present study we report on the detection of the HCNH$^+$ and HC$_3$NH$^+$ ions for the first time in a pre-stellar core. These ions are involved in the chemistry of HCN and HNC, both important species in the formation of larger and more complex nitrogen bearing molecules.}\\

%%%%%%%%%%%%%%%%%%%%%%%%%%%%%%%%%%%%%%%%%%%%%%%%%%

\section{Observations}

We used the observations published by \citet{vastel2014} combined with more recent observations. The latest were performed in December 2015 at the IRAM-30m toward L1544 ($\alpha_{2000} = 05^h04^m17.21^s, \delta_{2000} = 25\degr10\arcmin42.8\arcsec$). We used the broad-band receiver EMIR at 3mm, connected to an FTS spectrometer in its 50 kHz resolution mode. The beam of the observations is 33$''$, 28$''$ and 26$''$ at 75, 87 and 95 GHz, respectively. Weather conditions were very good with 2 to 3\,mm of precipitable water vapour. System temperatures were between 80 and 110\,K, resulting in an average rms of $\sim$4\,mK in a 50 kHz frequency bin. In order to obtain a flat baseline, observations were carried out using a nutating secondary mirror, with a throw of 3 arcmin. No contamination from the reference position was observed. Pointing was checked every 1.5\,hours on the nearby continuum sources 0439+360 and 0528+134. Pointing errors were always within 3$\arcsec$. We adopted the telescope and receiver parameters (main-beam efficiency, half power beam width, forward efficiency) from the values monitored at IRAM-30m (\url{http://www.iram.fr}). Line intensities are expressed in units of main-beam brightness temperature.

HCNH$^+$ contains a nitrogen atom and exhibits electric quadrupole hyperfine structure, resolved in our observations, whose frequencies and other spectroscopic parameters have been retrieved from the CDMS database\footnote{http://www.astro.uni-koeln.de/} based on astronomical observations of \citet{ziurys1992}. They have detected the electric quadrupole structure in this ion, through the observation of the J=1--0 transition toward the cold dark cloud TMC-1, enabling the determination of the quadrupole coupling constant.
The HCNH$^+$ rest frequencies from the CDMS database are the following: F\,=\,0--1: 74111.5427, F\,=\,2--1: 74111.3258, and F\,=\,1--1: 74111.1812\,MHz. The V$_{\textrm{LSR}}$ in L1544 is not constant due to gas motion and it could modify the frequency at which a given line is detected. We measured the V$_{\textrm{LSR}}$ (7.2$\pm$0.1\,km\,s$^{-1}$) for the strongest hyperfine structure (hereafter HFS) transition (F\,=\,2--1) of HCNH$^+$. The best fit for the F\,=\,0--1 transition is consistent with the CDMS frequency at 74111.54 MHz. The best fit for the F\,=\,1--1 transition (74111.14$\pm$0.01\,MHz) is offset by $\sim$0.04 MHz compared to the CDMS database (74111.1812\,MHz), within the uncertainty of 75\,kHz quoted from \citet{ziurys1992}.

{ In addition, we report observations of related species: HNC, HCN, HN$^{13}$C, H$^{13}$CN, and HC$_3$N. These species are important to understand the chemical origin of HCNH$^+$ and HC$_3$NH$^+$ and constrain better the chemical models, as discussed in following sections. The spectroscopic parameters for these species, as well as for HC$_3$NH$^+$, were also taken from the CDMS database. Unfortunately, HCN$^+$ and HNC$^+$ are not available in the databases but we are using HC$_3$N, HCN, and HNC to study the trends of the related species HCNH$^+$ and HC$_3$NH$^+$.}

Table \ref{lines} reports the spectroscopic parameters as well as the properties of the detected lines, obtained by Gaussian fitting, using the Levenberg-Marquardt { algorithm. For the line identification, we used the CASSIS\footnote{\url{http://cassis.irap.omp.eu}} software,} developed at IRAP \citep{vastel2015}. The detected spectra are shown in Fig. \ref{hcnhp_lines}{ , \ref{hc3nhp_lines}, \ref{hc3n_lines}, and \ref{hcn_hnc_lines}}.

\begin{table*} 
%\tiny
\centering
\caption{{ Properties of the observed HC$_3$NH$^+$, HCNH$^+$, HCN, HNC, H$^{13}$CN, HN$^{13}$C, and HC$_3$N lines (E$_{\textrm{up}}$ $\le$ 35\,K, A$_{\textrm{ul}}$ $\ge$ 10$^{-7}$\,s$^{-1}$). Uncertainties are given at the 1 sigma level and only include statistical fluctuations of the Gaussian fits of the lines. The rms has been computed over 20\,km\,s$^{-1}$. The dash symbol means that the line show self absorption, so not fit has been performed. \label{lines}}}
\begin{tabular}{ccccccccc}
\hline\hline
Transitions	&Frequency	&E$_{\textrm{up}}$	&A$_{\textrm{ul}}$	&${\rm S\mu^2}$	&V$_{\textrm{peak}}$	&FWHM			&$\int T_{\textrm{mb}}dV$		&rms\\
(J',F'--J,F)		&(MHz)		&(K)				&(s$^{-1}$)		&				&(km\,s$^{-1}$)			&(km\,s$^{-1}$)		&(mK\,km\,s$^{-1}$)			&mK\\
\hline
&&&& HCNH$^+$ &&&&\\
\hline
1,0 -- 0,1  & 74111.54  & 3.56  &1.33 10$^{-7}$ & 0.028 &   &   & 11 $\pm$ 3 &  \\
1,2 -- 0,1  & 74111.33 & 3.56  &1.33 10$^{-7}$ & 0.14 &  7.20 $\pm$ 0.10 & 0.39 $\pm$ 0.11  & 42 $\pm$ 8  & 4.3\\
1,1 -- 0,1  & 74111.14 & 3.56  &1.33 10$^{-7}$ & 0.084 & &   & 27 $\pm$ 3 &\\
\hline
&&&& HC$_3$NH$^+$ &&&&\\
\hline
9,8 -- 8,7		& 77920.63  & 18.70  &6.67 10$^{-6}$& 20.5873 & 7.22 $\pm$ 0.05  & 0.37 $\pm$ 0.07  & 6.4 $\pm$ 2.1 & 3.7 \\
9,9 -- 8,8		& 77920.63 & 18.70  &6.68 10$^{-6}$ & 23.0409 &  &   &  &\\
9,10 -- 8,9  	& 77920.63 & 18.70  &6.76 10$^{-6}$ & 25.7870 &  &   &  &\\
10,9 -- 9,8  	& 86578.14  & 22.85  &9.22 10$^{-6}$ & 23.1953 & 7.20 $\pm$ 0.05  & 0.41 $\pm$ 0.07  & 5.1 $\pm$ 1.6 & 3.6 \\
10,10 -- 9,9  	& 86578.14 & 22.85  &9.23 10$^{-6}$ & 25.6626 &  &   &  &\\
10,11 -- 9,10 	& 86578.14 & 22.85  &9.23 10$^{-6}$ & 28.3923 &  &   &  &\\
11,10 -- 10,9 	& 95235.53  & 27.42  &1.24 10$^{-5}$ & 25.7988 & 7.26 $\pm$ 0.05  & 0.63 $\pm$ 0.05  & 12.0 $\pm$ 2.3 & 3.8 \\
11,11 -- 10,10  	& 95235.53 & 27.42  &1.24 10$^{-5}$ & 28.2748 &  &   &  &\\
11,12 -- 10,11 	& 95235.53 & 27.42  &1.25 10$^{-5}$ & 30.9885 &  &   &  &\\
\hline
&&&& HC$_3$N &&&&\\
\hline
8 -- 7       & 72783.82   & 15.72   & 2.94 10$^{-5}$ & 111.429 & 7.23 $\pm$ 0.01 & 0.51 $\pm$ 0.01   & 2013 $\pm$ 5 & 5.1\\
9 -- 8       & 81881.47   & 19.65   & 4.21 10$^{-5}$ & 125.314 & 7.23 $\pm$ 0.01 & 0.54 $\pm$ 0.01   & 1610 $\pm$ 3 & 3.7\\
10 -- 9     & 90979.02   & 24.02   & 5.81 10$^{-5}$ & 139.252 & 7.33 $\pm$ 0.05 & 0.50 $\pm$ 0.03   & 1367 $\pm$ 2 & 1.7\\
11 -- 10   & 100076.39 & 28.82   & 7.77 10$^{-5}$ & 153.179 & 7.24 $\pm$ 0.02 & 0.49 $\pm$ 0.03   & 939 $\pm$ 2 & 3.9\\
12 -- 11   & 109173.63 & 34.06   & 1.04 10$^{-4}$ & 167.109 & 7.23 $\pm$ 0.01 & 0.46 $\pm$ 0.03   & 617 $\pm$ 2 & 3.2\\
\hline
&&&& HCN &&&&\\
\hline
1,1 -- 0,1       & 88630.42   & 4.25   & 2.41 10$^{-5}$ & 8.922 & - & -   & 698 $\pm$ 5 & 5.1\\
1,2 -- 0,1       & 88631.85   & 4.25   & 2.41 10$^{-5}$ & 14.869 & - & -   & 1083 $\pm$ 5 & 5.1\\
1,0 -- 0,1       & 88633.94   & 4.25   & 2.41 10$^{-5}$ & 2.974 & - & -   & 802 $\pm$ 5 & 5.1\\
\hline
&&&& H$^{13}$CN &&&&\\
\hline
1,1 -- 0,1       & 86338.73   & 4.14   & 2.22 10$^{-5}$ & 8.890 & 7.14 $\pm$ 0.01 & 0.52 $\pm$ 0.02   & 209 $\pm$ 2 & 4.0\\
1,2 -- 0,1       & 86340.16   & 4.14   & 2.22 10$^{-5}$ & 14.883 & 7.12 $\pm$ 0.01 & 0.56 $\pm$ 0.02   & 303 $\pm$ 2 & 4.0\\
1,1 -- 0,1       &  86342.25   & 4.14   & 2.23 10$^{-5}$ & 2.976 & 7.17 $\pm$ 0.01 & 0.51 $\pm$ 0.01   & 112 $\pm$ 2 & 4.0\\
\hline
&&&& HNC &&&&\\
\hline
1 -- 0       & 90663.57   & 4.35   & 2.69 10$^{-5}$ & 9.302 & - & -   & 2384 $\pm$ 5 & 4.3\\
\hline
&&&& HN$^{13}$C &&&&\\
\hline
1 -- 0       & 87090.83   & 4.18   & 2.38 10$^{-5}$ & 9.302 & 7.16 $\pm$ 0.03 & 0.82 $\pm$ 0.05   & 1212 $\pm$ 5 & 4.1\\
\hline
\end{tabular}
\end{table*} 

\begin{figure}
   \centering
   \includegraphics[width=\hsize,clip=true,trim=0 0 0 0]{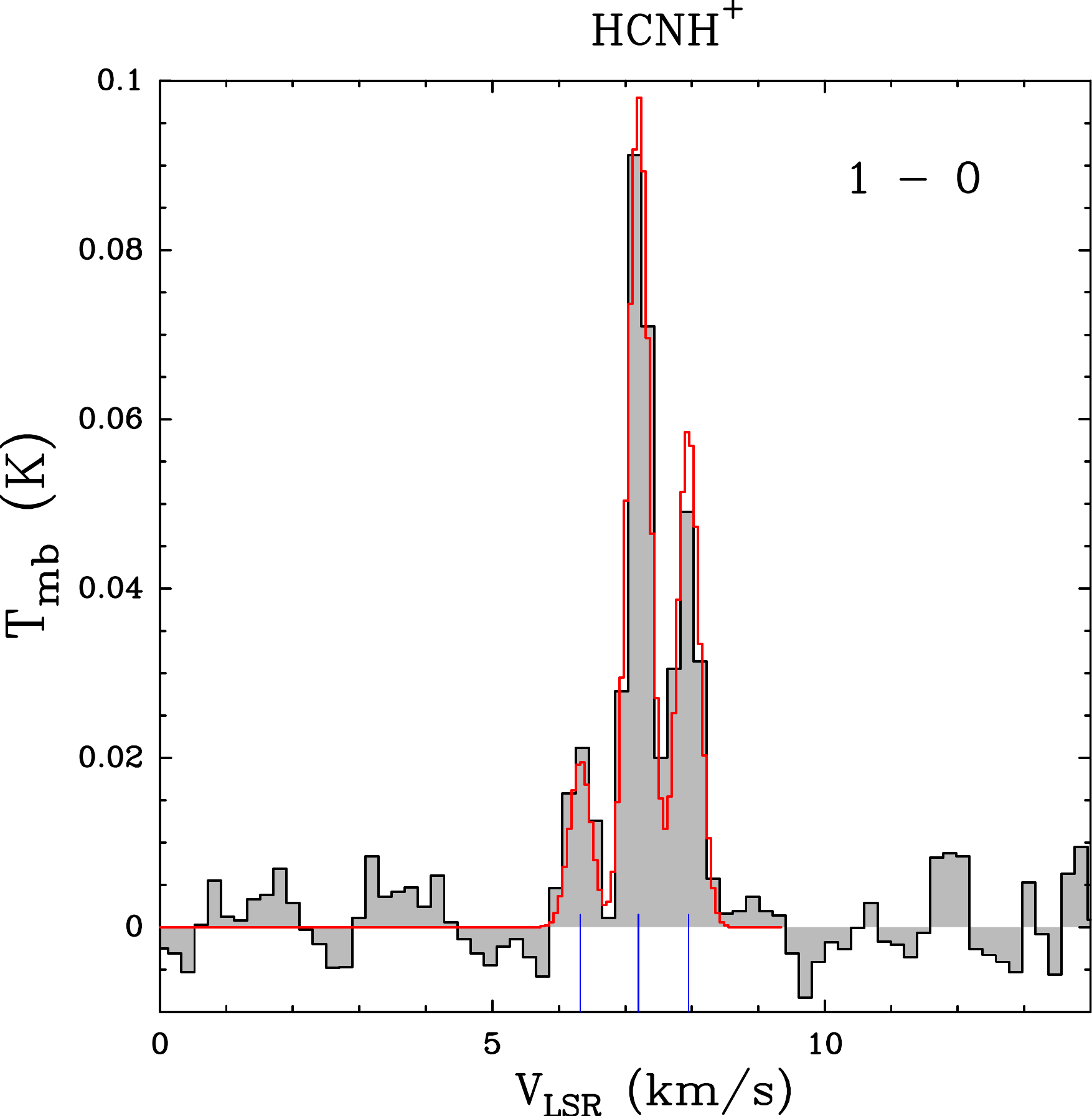}
   \caption{Observed spectra of the HCNH$^+$ 1--0 HFS transitions at 74.1\,GHz. Temperatures are in main beam. The red line shows the LTE fit at 10 K using the following frequencies: F\,=\,0--1 at 74111.54 MHz, F\,=\,{ 2--1} at 74111.33 MHz, F\,=\,1--1 at 74111.14 MHz. Intensities are given in units of main beam temperature in K.}
   \label{hcnhp_lines}
 \end{figure}

 \begin{figure}
   \centering
   \includegraphics[width=\hsize,clip=true,trim=0 0 20 0]{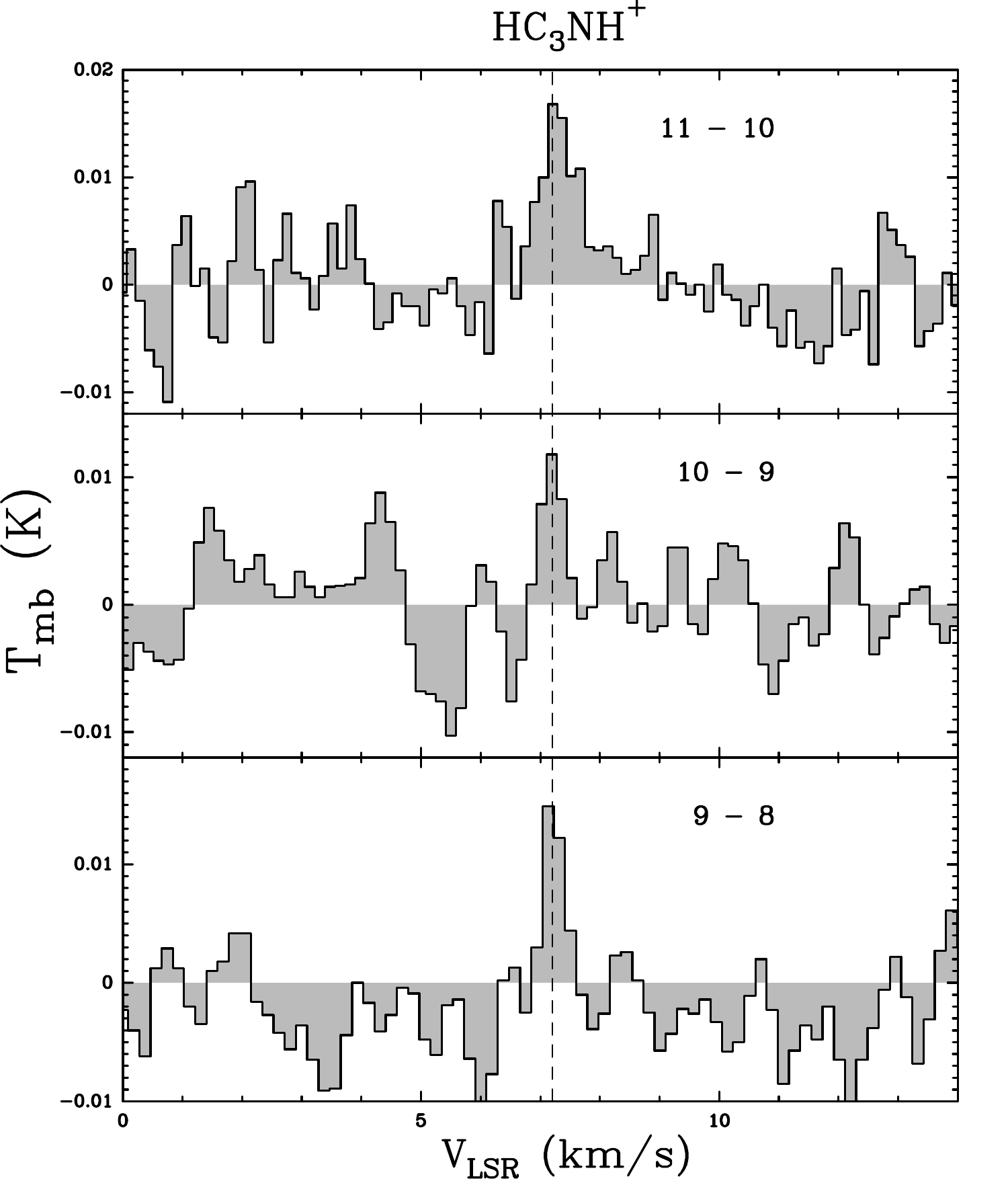}
   \caption{Observed spectra of the HC$_3$NH$^+$ 9--8 (bottom), 10--9 (middle), and 11--10 (up) transitions at 77.9\,GHz, 86.8\,GHz, and 95.2\,GHz, respectively. The dashed line shows a V$_{\textrm{LSR}}$ of 7.2\,km\,s$^{-1}$. Intensities are given in units of main beam temperature in K.}
   \label{hc3nhp_lines}
 \end{figure}

  \begin{figure}
   \centering
   \includegraphics[width=\hsize,clip=true,trim=0 20 20 0]{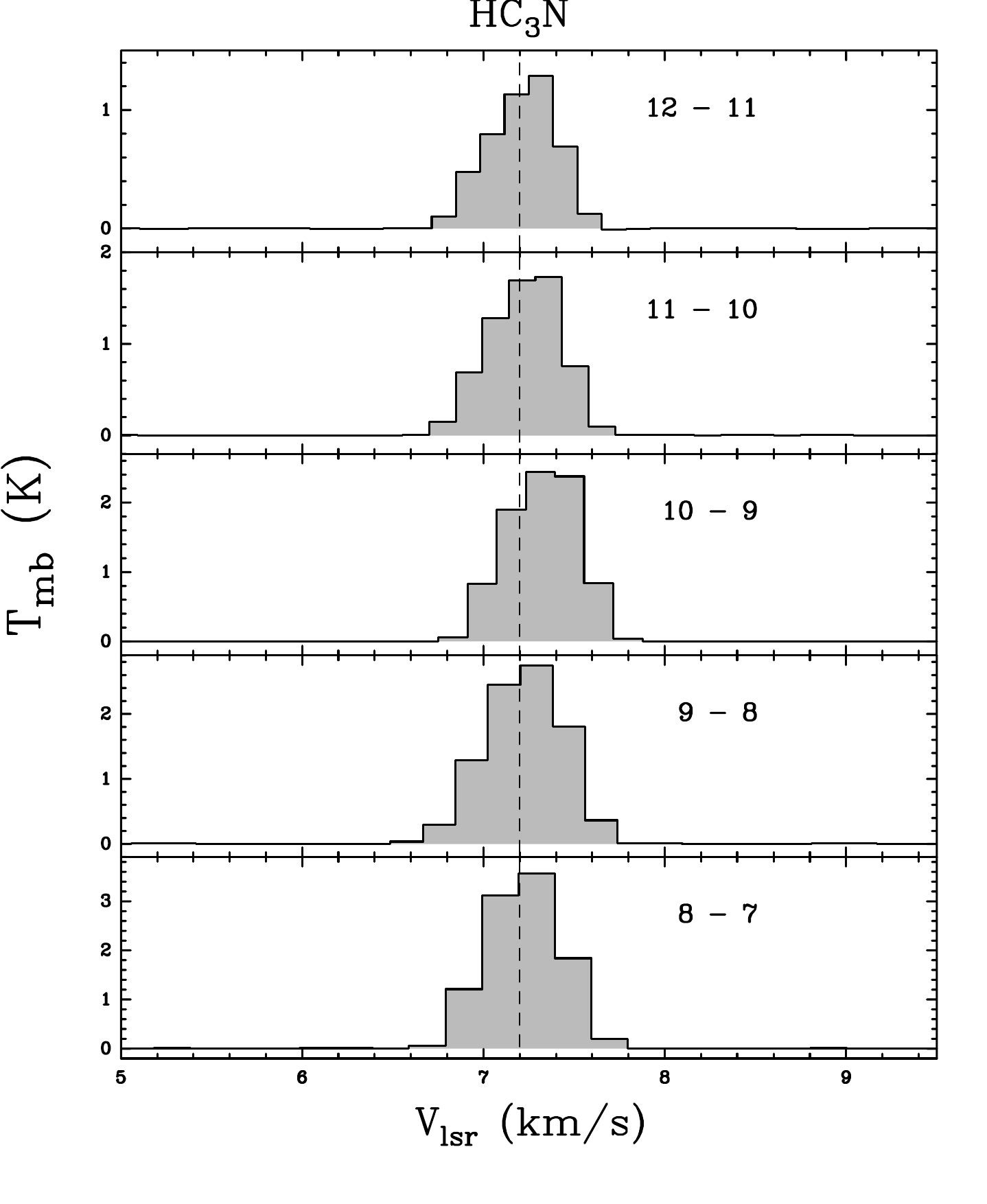}
   \caption{{ Observed spectra of the HC$_3$N transitions at 72.8\,GHz (8--7), 81.9\,GHz (9--8), 91\,GHz (10--9), 100.1\,GHz (11--10), 109.2\,GHz (12--11). The dashed line shows a V$_{\textrm{LSR}}$ of 7.2\,km\,s$^{-1}$. Intensities are given in units of main beam temperature in K.}}
   \label{hc3n_lines}
 \end{figure}
 
 \begin{figure}
   \centering
   \includegraphics[width=\hsize,clip=true,trim=0 0 0 0]{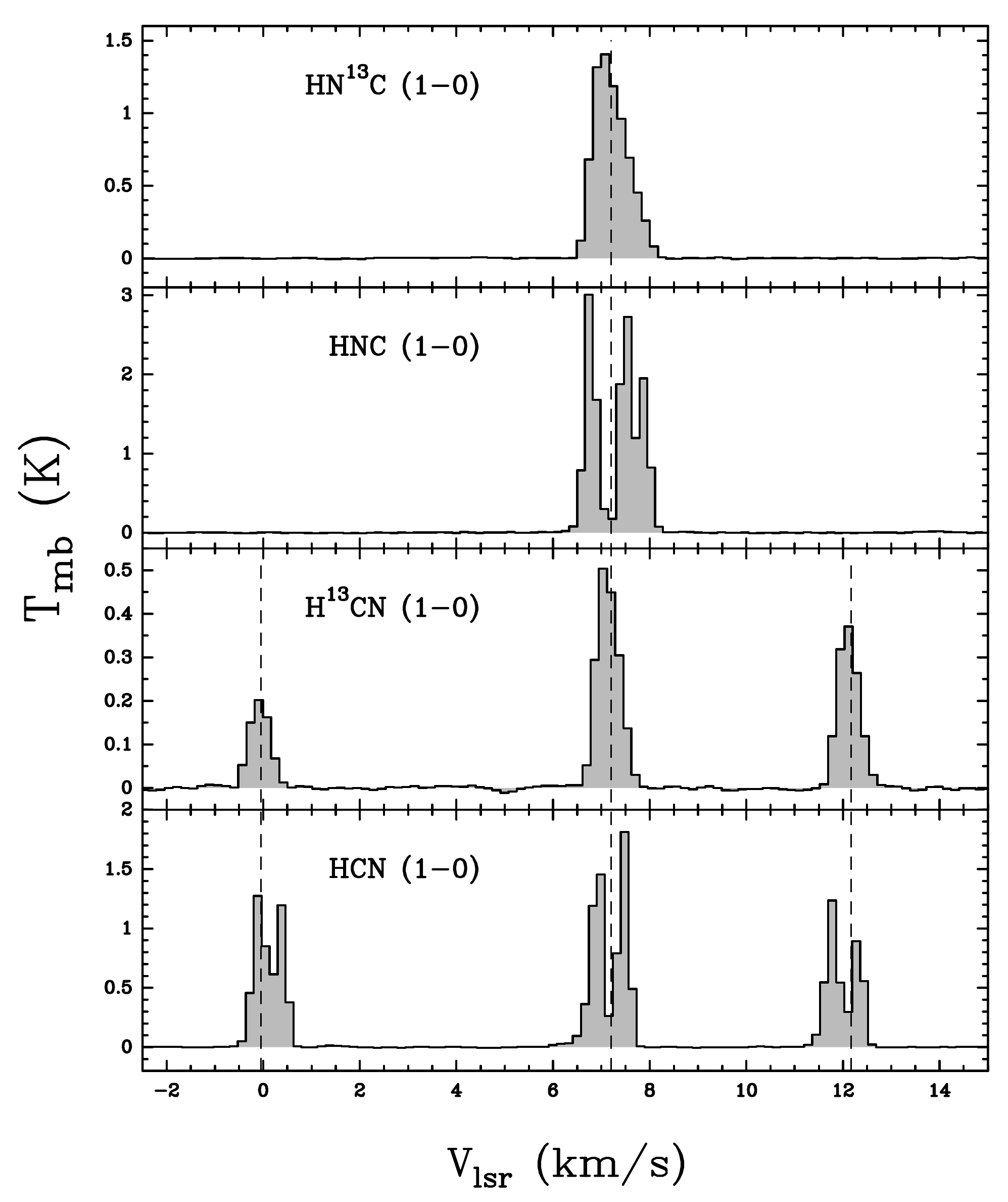}
   \caption{ Observed spectra of the HCN, H$^{13}$CN, HNC and HN$^{13}$C 1--0 transition at 88.6, 86.3, 90.7 and 87.1 GHz respectively. Intensities are given in units of main beam temperature in K. The dashed line shows a V$_{\textrm{LSR}}$ of 7.2\,km\,s$^{-1}$ for both HNC and HN$^{13}$C and indicates the same velocity for the hyperfine structure of the HCN and  H$^{13}$CN 1--0 transition.}
   \label{hcn_hnc_lines}
 \end{figure}

%%%%%%%%%%%%%%%%%%%%%%%%%%%%%%%%%
\section{Results}\label{results}

{ We performed an LTE analysis of the observed line intensities for HCNH$^+$, HC$_3$NH$^+$, and HC$_3$N using CASSIS, assuming no beam dilution. Note that it is consistent with the chemical modelling performed in Section \ref{chem_mod}. Considering the low temperatures in the L1544 pre-stellar core, we used a 10\,K excitation temperature for the detected transition of HCNH$^+$ in order to determine its column density: N(HCNH$^+$)\,=\,(2.0$\pm$0.2)$\times$10$^{13}$\,cm$^{-2}$. From the HC$_3$NH$^+$ (9--8), (10--9), and (11--10) transitions, the excitation temperature must be in the [10\,K\,--\,15\,K] range with a column density N(HC$_3$NH$^+$)\,=\,[1--2]$\times10^{11}$\,cm$^{-2}$ to reproduce the observations. We checked that a change in the excitation temperature in the range [5\,K\,--\,15\,K] does not change the resulting column densities of HCNH$^+$ and HC$_3$NH$^+$ by more than a factor of $\sim$1.4. Fig. \ref{hcnhp_lines} shows the LTE model with an excitation temperature of 10\,K for the HCNH$^+$ transition over-plotted with the observations. All lines are optically thin.

For HC$_3$N, we carried out a non-LTE analysis using the LVG code by \citet{ceccarelli2003}. For that, we used the collision coefficients with H$_2$ \citep{faure2016} and derived a kinetic temperature of $12\pm2$\,K and a density of (5$\pm$3)$\times$10$^4$\,cm$^{-3}$ with a column density N(HC$_3$N)\,=\,[1.6--2.4]$\times10^{13}$\,cm$^{-2}$. We also performed an LTE analysis to compare with the non-LTE result. In order to fit simultaneously all five HC$_3$N transitions, the excitation temperature must be T$_{\textrm{ex}}$\,=\,8.5$\pm$0.5\,K with N(HC$_3$N)\,=\,(1.9$\pm$0.2)$\times10^{13}$\,cm$^{-2}$. The LTE and non-LTE results, therefore, give similar column densities.

We present on Fig. \ref{hcn_hnc_lines} the HCN, H$^{13}$CN, HNC and HN$^{13}$C (1--0) transitions obtained with our ASAI observations. The HCN and HNC line profiles show self-absorption and so-called hyperfine anomalies \citep{guilloteau1981,gonzalez-alfonso1993}, indicating opacity effect. Therefore, to determine the HCN and HNC column densities, we used their respective isotopologues. For H$^{13}$CN we used the collision rates from \citet{green1974-1} with Helium, scaled by a factor of 1.363 to reproduce the collisions by H$_2$. The resulting HCN column density is [2.7--4.4]$\times10^{14}$\,cm$^{-2}$ using a $^{12}$C/$^{13}$C\,=\,68, for a kinetic temperature between 8 and 10 K and a density between 5$\times$10$^4$ and 10$^5$\,cm$^{-3}$. The H$^{13}$CN line is optically thick and the column density is therefore uncertain. For HN$^{13}$C we used the collision rates from \citet{dumouchel2011}. The resulting HNC column density is [2--4]$\times$10$^{14}$\,cm$^{-2}$ using a $^{12}$C/$^{13}$C\,=\,68, for a kinetic temperature between 8 and 10 K and a density between 5$\times$10$^4$ and 10$^5$ cm$^{-3}$. The HN$^{13}$C is moderately optically thick with a low value for the excitation temperature at 4 K. Such low excitation temperatures were also reported by \citet{padovani2011}. Both HCN and HNC have previously been observed by \citet{hily-blant2010} using the IRAM 30\,m telescope with a spectral resolution of 20\,kHz of the J=1--0 HN$^{13}$C and H$^{13}$CN species. Assuming T$_{\textrm{ex}}$\,=\,8\,K and a $^{12}$C/$^{13}$C\,=\,68, they derived N(HCN)\,=\,[1.9--2.9]$\times10^{13}$\,cm$^{-2}$ and N(HNC)\,=\,[3.8--5.8]$\times10^{13}$\,cm$^{-2}$ leading to a HNC/HCN ratio of $\sim$\,2.0. The resulting excitation temperature from our analysis is much lower than the 8 K value assumed by \citet{hily-blant2010}. We note that lowering their excitation temperature to 3.5--4 K (instead of their assumed value of 8\,K) would increase their HCN column density by a factor of 5. We use the results from our observations and analysis in the present work.

{We also present the observed column densities for HCO$^+$, N$_2$H$^+$ (two other relevant ions in this pre-stellar core) and CO. These three species will allow us to test the completeness and robustness of the chemical study. To perform the comparison with the modellings, we have taken the observations derived by \citet{caselli2002-1} and \citet{caselli2002}. In their study, the authors obtained N(HCO$^+$)\,=\,[0.3--1.0]$\times10^{14}$\,cm$^{-2}$ (T$_{\textrm{ex}}$\,$\sim$\,8\,K), N(N$_2$H$^+$)\,=\,[0.3--2.0]$\times10^{13}$\,cm$^{-2}$ (T$_{\textrm{ex}}$\,=\,5\,K), and N(CO)\,=\,[0.6--1.7]$\times10^{18}$\,cm$^{-2}$ (T$_{\textrm{ex}}$\,=\,10\,K).}
}

%%%%%%%%%%%%
\section{Chemical modelling}\label{chem_mod}

\subsection{Description of the modelling}

{ In this section, we aim at deriving the spatial variation of the abundance of HCNH$^+$ and HC$_3$NH$^+$ for several chemical ages of the pre-stellar core. We do not pretend here to describe the situation with a self-consistent model, but just to shed light on the formation pathways of both ionic species HC$_3$NH$^+$ and HCNH$^+$. We will confront the predicted column densities with the observed values.}

To do so, we have used the gas-grain chemical code \textit{Nautilus} \citep{hersant2009, ruaud2016} to predict the abundances of both species in the cold core. \textit{Nautilus} allows to compute the evolution of the chemical composition of the gas and the icy mantle of the grains.
Details on the processes included in the model can be found in \citet{ruaud2016}. Note that we have used \textit{Nautilus} in its two-phase model, meaning that there is no distinction between the surface and the bulk of the mantle of the grains.
{ Briefly, gas species can stick on the grain mantles where they can diffuse (depending on the species), be photo-dissociated and undergo chemical reactions. In addition, frozen species are injected into the gas phase by thermal and non thermal processes. For the latter, we considered: (i) photo-desorption by direct and cosmic-ray induced UV photons, with a fixed yield equal to $10^{-4}$; (ii) direct cosmic-ray desorption, following the formalism by \citet{hasegawa1993}; (iii) chemical desorption, following the formalism by \citet{garrod2007}, in which a fixed 1\% of the species formed on the grain surfaces is injected in the gas-phase, regardless the species.}
The gas phase reactions are based on the kida.uva.2014 network\footnote{\url{http://kida.obs.u-bordeaux1.fr/networks.html}} \citep[see][]{wakelam2015} while the surface network is based on \citet{garrod2006}. The full network contains 736 species (488 in the gas phase and 248 at the surface of the grains) and 10466 reactions (7552 pure gas phase reactions and 2914 reactions of interactions with grains and reactions at the surface of the grains).

\begin{table} 
	\centering
	\caption{ Initial gas phase elemental abundances assumed relative to the total nuclear hydrogen density n$_{\textrm{H}}$.\label{EAmodels}}
	\begin{tabular}{lcc}
		\hline\hline
		Species	&	EA1	&	EA2\\
		\hline
		He		&	$1.40\times10^{-1}$	&	$9.00\times10^{-2}$\\
		N		&	$2.14\times10^{-5}$	&	$7.60\times10^{-5}$\\
		O		&	$1.76\times10^{-4}$	&	$2.56\times10^{-4}$\\
		C$^+$	&	$7.30\times10^{-5}$	&	$1.20\times10^{-4}$\\
		S$^+$	&	$8.00\times10^{-8}$	&	$1.50\times10^{-5}$\\
		Si$^+$	&	$8.00\times10^{-9}$	&	$1.70\times10^{-6}$\\
		Fe$^+$	&	$3.00\times10^{-9}$	&	$2.00\times10^{-7}$\\
		Na$^+$	&	$2.00\times10^{-9}$	&	$2.00\times10^{-7}$\\
		Mg$^+$	&	$7.00\times10^{-9}$	&	$2.40\times10^{-6}$\\
		Cl$^+$	&	$1.00\times10^{-9}$	&	$1.80\times10^{-7}$\\
		P$^+$	&	$2.00\times10^{-10}$&	$1.17\times10^{-7}$\\
		F$^+$	&	$6.68\times10^{-9}$	&	$1.80\times10^{-8}$\\
		\hline
	\end{tabular}
\end{table}

The chemical model is divided into two phases. { The first phase corresponds to the evolution of the chemistry in a diffuse or molecular cloud (hereafter ambient cloud), depending on the initial H density assumed (see below). We let the chemical composition evolve until $1\times10^6$ years. The abundances from this first step are then used as initial abundances for the second step where we consider the physical structure of the L1544 pre-stellar core, as determined by \citet{keto2014}.\\

\textit{Ambient cloud phase:} The first phase takes into account photo-processes, for both the gas and grain surface chemistry. We adopted two different sets of initial atomic abundances (with respect to the total proton density n$_\textrm{H}$) given in Table \ref{EAmodels} \citep{wakelam2008}:}
\begin{enumerate}
	\item In model EA1, which is consistent with recent studies performed towards this core \citep{vasyunin2013, jimenez-serra2016}, initial abundances are called ``low-metal abundances'' because they are lower by a factor of $\sim$100 for heavy elements in comparison to solar elemental abundances in order to take into account elemental depletion on grains in cold interstellar cores. The carbon and oxygen abundances are respectively $7.30\times10^{-5}$ and $1.76\times10^{-4}$ leading to a C/O ratio of $\sim$0.4.
	\item The model EA2, based on the high-metal abundances observed in the $\zeta$ Oph diffuse cloud, but modified based on recent observations. The carbon and oxygen abundances are respectively $1.20\times10^{-4}$ and $2.56\times10^{-4}$ leading to a C/O ratio of $\sim$0.5.
\end{enumerate}
For both sets, the C/O { ratio ($\sim$\,0.4 and $\sim$\,0.5 for models EA1 and EA2, respectively)} is consistent with recent results towards L1544 \citep{vastel2014}. Nonetheless, { by varying the initial oxygen abundance shown in Table \ref{EAmodels},} we have varied the C/O ratio to be 0.4, 0.5, 0.6, 0.9, and 1.2 but it does not change significantly the result. Therefore, in the following, we use the { initial gas phase elemental abundance shown in Table \ref{EAmodels} for a given model}. The gas and grain temperature are fixed to 20\,K and 10\,K respectively. The cosmic ray ionisation rate is set to the standard value of $1.3\times10^{-17}$\,s$^{-1}$. These values are consistent with recent studies of this core \citep[e.g.][]{vastel2016, jimenez-serra2016}. We have verified that a difference of 10\,K for both the gas or grain temperatures will not change the result significantly. In the literature, various initial H densities have been used to describe this { ambient cloud} phase, ranging from n$_\textrm{H}=10^2$ to a few $10^4$\,cm$^{-3}$ \citep[][in order of increasing initial H densities assumed in their studies]{jimenez-serra2016, quenard2016, vastel2016}. One must also note that different chemical codes have been used in these different studies. For the sake of our study, we have tested different initial H densities in the range defined above (see Table \ref{init_dens}). The assumed size of the ambient cloud is $\sim$6$\times10^4$\,au or 0.3\,pc \citep{quenard2016}. This leads to $A_V$ up to $\sim$10\,mag depending on the initial H density taken.\\

\begin{table}
	\centering
	\caption{Chemical modellings parameters shown in this study.\label{init_dens}}
	\begin{tabular}{llc}
	\hline\hline
	Model 	&	Density~$^a$	&	Elemental\\
	number	&	(cm$^{-3}$)	&	abundances model\\
	\hline
	1	&	$1\times10^2$~$^b$	&	EA1\\
	2	&	$3\times10^3$~$^c$	&	EA1\\
	3	&	$2\times10^4$~$^d$	&	EA1\\
	4	&	$1\times10^2$	&	EA2\\
	5	&	$3\times10^3$	&	EA2\\
	6	&	$2\times10^4$	&	EA2\\
	\hline
	\multicolumn{3}{l}{$^a$ Initial H densities used in the ambient}\\
	\multicolumn{3}{l}{cloud phase.}\\
	\multicolumn{3}{l}{$^b$ Value used in \citet{vasyunin2013}}\\
	\multicolumn{3}{l}{and \citet{jimenez-serra2016}.}\\
	\multicolumn{3}{l}{$^c$ Value used in \citet{quenard2016}.}\\
	\multicolumn{3}{l}{$^d$ Value used in \citet{vastel2014} and}\\
	\multicolumn{3}{l}{\citet{vastel2016}.}
	\end{tabular}
\end{table}

{ \textit{Pre-stellar phase:}} For the second phase, the final abundances of the first phase are used as initial abundances. The physical structure adopted in this phase (density, gas and grain temperature) is the one derived from the recent study of \citet{keto2014} { (see Fig. \ref{struct_l1544})}. This physical model has been widely used in recent studies to describe the structure of L1544 \citep[e.g.][]{jimenez-serra2016, quenard2016, vastel2016}. The cosmic ray ionisation rate is kept the same during this step. { In this second phase, the chemistry is followed during 3 millions years. At the end, the radial abundance profiles of the species for several ages of the pre-stellar core is obtained.
At the end of phase 2, carbon is largely depleted into the grain mantles in the innermost part of the condensation: at $5\times10^3$\,au, where the density is about $\sim$\,7$\times 10^4$\,cm$^{-3}$, only 0.1\% of carbon is in the gas phase (mainly in the form of CO), the rest is almost equally distributed into iced CO, H$_2$CO and CH$_3$OH. At about $10^4$\,au the gas phase carbon amounts to about 1\%, while in the very external regions, where UV photons penetrate, gaseous carbon varies from 10 to 90\% for the density varying from $2\times10^4$ to $1\times10^2$\,cm$^{-3}$. Similar values are also found for gaseous oxygen and nitrogen. At about $10^4$\,au, the most abundant gaseous O-bearing species are CO and the atomic O and the N-bearing species are N$_2$ and atomic N, whereas the most abundant iced species are H$_2$O and NH$_3$, respectively.}\\

\begin{figure}
	\centering
	\includegraphics[width=\hsize,clip=true,trim=0 0 0 0]{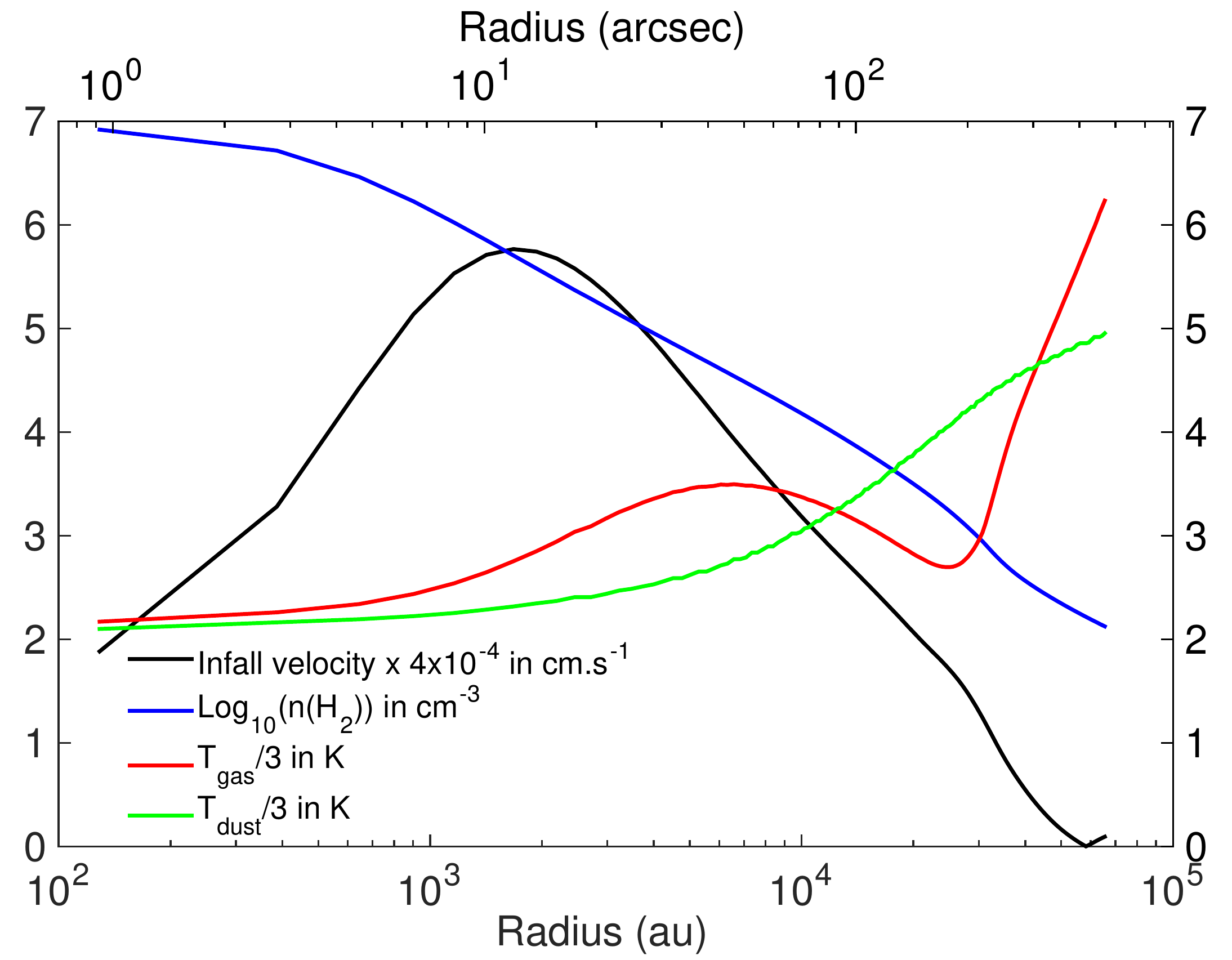}
	\caption{{ Gas and dust temperature, density, and velocity profiles of L1544 used in this study from \citet{keto2014}.}}
	\label{struct_l1544}
\end{figure}

\subsection{Results and comparison with observations}

{ We have considered six different chemical models using the three different initial gas densities of the ambient cloud phase presented previously and the two different sets (EA1 and EA2) of initial atomic abundances (see models shown in Table \ref{init_dens}).}
To easily compare the modelled abundance with the observations, we calculated the column density along the radius of the core. We used the method described in \citet{jimenez-serra2016} to convert the abundance $[X]$ (with respect to H) of a species $X$ into column densities $N(X)$:
\begin{equation}\label{eq_Ncol}
	N(X) = 2\,\times\,\sum_{i=2}^{n}\left(\frac{n(\textrm{H})_i[X]_i + n(\textrm{H})_{i-1}[X]_{i-1}}{2}\right)\,\times\,(R_{i-1} - R_{i})
\end{equation}
with $R$ the radius from the centre and $i$ the position in the grid along the line of sight ($i=1$ being the outermost position) composed of $n$ shells ($n=129$). $n(\textrm{H})_i$ is the gas density at radial point $i$, $[X]_i$ the abundance of the species. Different column densities are derived depending on the observed position towards the core.
The different column densities obtained using Eq. (\ref{eq_Ncol}) are then} convolved with the beam size of the IRAM-30m telescope at the frequency of the observations ($\sim$30$''$) to take into account this effect. The H$_2$ column density has also been calculated following the same method and we derive N(H$_2$)\,=\,$6.7\times10^{22}$\,cm$^{-2}$. Thus, the observed abundances (with respect to H$_2$) are [HCNH$^+$]$\,\simeq3\times10^{-10}$ and [HC$_3$NH$^+$]$\,\simeq[1.5-3.0]\times10^{-12}$.

\begin{figure*}
	\centering
	\includegraphics[width=0.49\hsize,clip=true,trim=0 0 0 0]{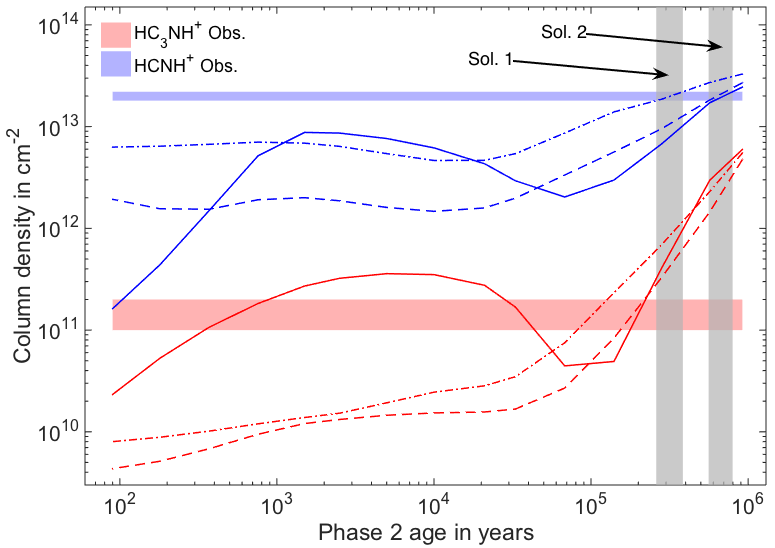}
	\includegraphics[width=0.49\hsize,clip=true,trim=0 0 0 0]{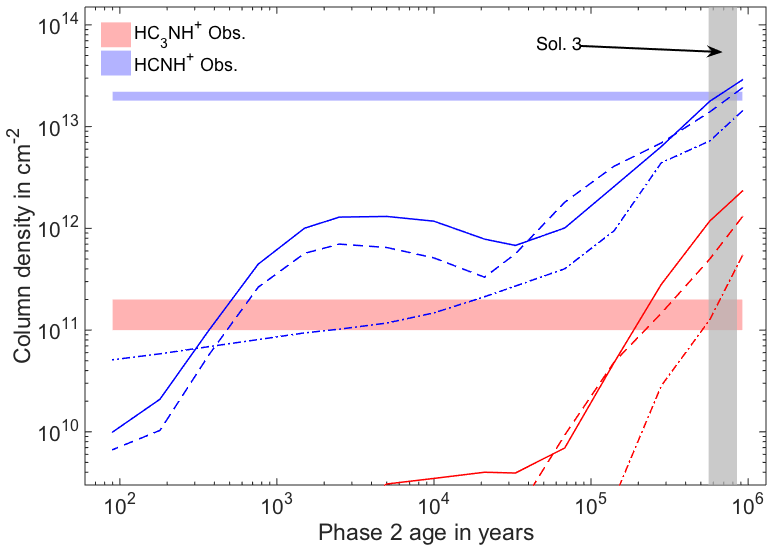}\\
	\includegraphics[width=0.49\hsize,clip=true,trim=0 0 0 0]{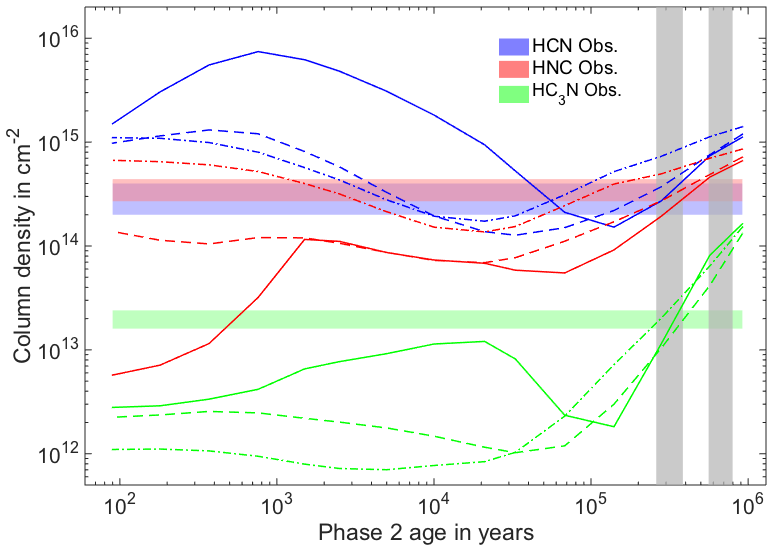}
	\includegraphics[width=0.49\hsize,clip=true,trim=0 0 0 0]{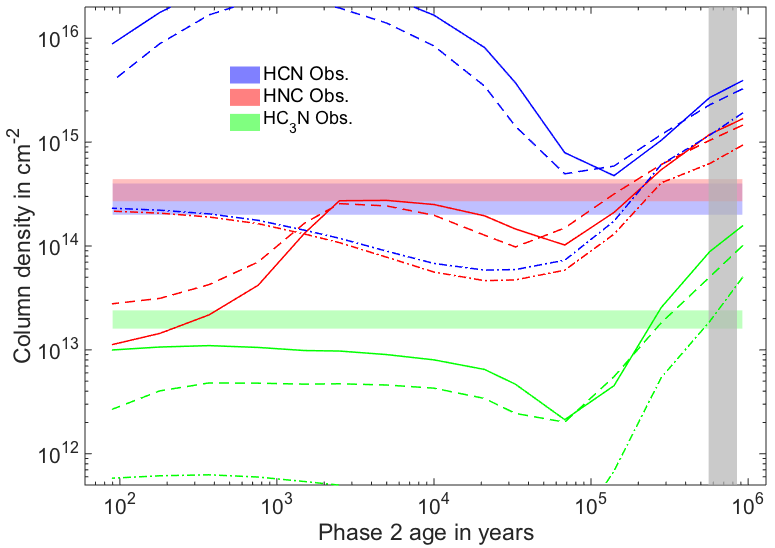}\\
	\includegraphics[width=0.49\hsize,clip=true,trim=0 0 0 0]{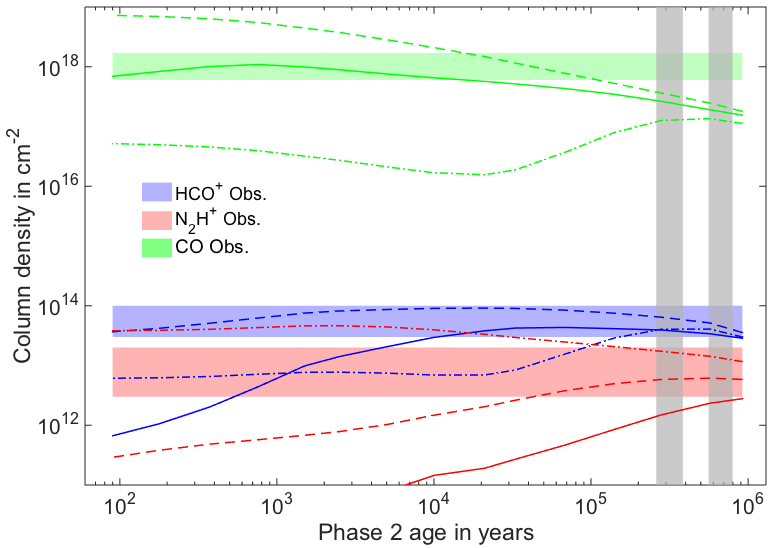}
	\includegraphics[width=0.49\hsize,clip=true,trim=0 0 0 0]{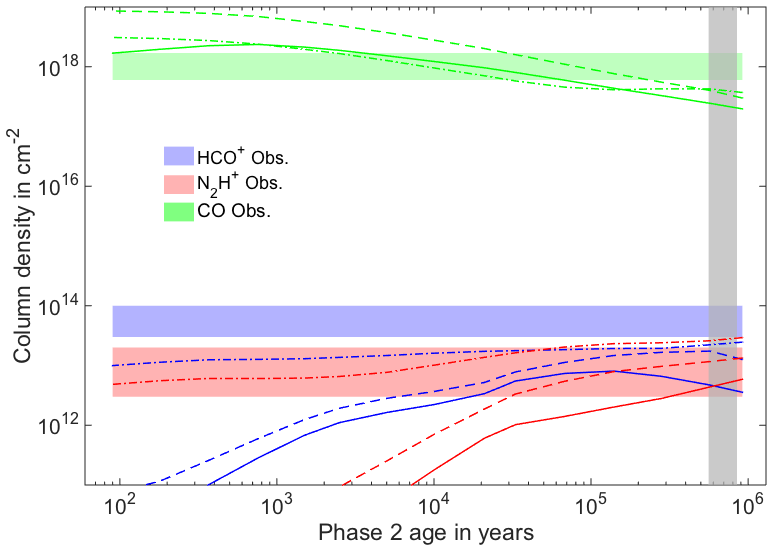}\\	
	\caption{ Column densities for HCNH$^+$, HC$_3$NH$^+$ (top panels), HCN, HNC, HC$_3$N (middle panels), HCO$^+$, N$_2$H$^+$, and CO (bottom panels) as a function of the age of the phase 2 for different initial gas densities of the ambient cloud phase: $1\times10^2$\,cm$^{-3}$ (full line), $3\times10^3$\,cm$^{-3}$ (dashed line), and $2\times10^4$\,cm$^{-3}$ (dash-dotted line). The area of confidence of the observed column densities for these species is also shown. Grey areas shows the timespan area of confidence of each model based on the observed HCNH$^+$. {These areas are labelled solutions 1, 2 and 3.} \textit{Left panels:} Model EA1 of initial atomic abundances. \textit{Right panels:} Model EA2 of initial atomic abundances.}
	\label{result_nautilus}
\end{figure*}

Figure \ref{result_nautilus} presents the different column densities { (HCNH$^+$, HC$_3$NH$^+$, HCN, HNC, HC$_3$N, HCO$^+$, N$_2$H$^+$, and CO) for models reported in Table \ref{init_dens}. Based on the observed HCNH$^+$, we have defined three vertical grey areas corresponding to three possible solutions for the age of phase 2 (solutions 1 and 2 for models 1, 2, and 3 and solution 3 for model 4, 5, and 6).} No perfect agreement between the models and the observations is found { to reproduce simultaneously both HCNH$^+$ and HC$_3$NH$^+$} for the different ages and different initial H densities. Nonetheless, above an age of $1\times10^5$\,yrs, both HCNH$^+$ or HC$_3$NH$^+$ column density predictions cross the observed column density, unfortunately not at the same age. Either HCNH$^+$ is underestimated or HC$_3$NH$^+$ is overestimated, by a factor of $\sim$3--15, depending on the initial H density. 

{ The middle panels of Figure \ref{result_nautilus} present the column densities of HCN, HNC, and HC$_3$N. The observed column density of HC$_3$N (green area) is reproduced by all three different initial densities of models 1, 2 and 3 within the same timescale of solution 1 (shown in grey area). It is also reproduced with model 6 for solution 3. No models can reproduce HC$_3$N for solution 2. The observed HNC column density (red area) is reproduced with models 1 and 2 and within a factor of $\sim$3 for model 3. Solutions 2 reproduces the observed HNC within a factor of $\sim$3. For solution 3, the model 6 is the only one reproducing the observed HNC column density below a factor of $\sim$3. For HCN (in blue), we manage to reproduce the observed column density with solution 1 (model 1 and 2) and within a factor of $\sim$3 for model 3. For solutions 2 and 3, we obtain a modelling-to-observed ratio $\lesssim$5, and $\lesssim$10 (respectively) for this species. Around 3$\times$10$^5$\,yrs, for all models of solution 1, HC$_3$N, HCN and HNC are well reproduced. At the same age, for any of these models, HCNH$^+$ and HC$_3$NH$^+$ tend to be under-estimated and over-estimated (respectively). Since these ionic and neutral species are closely related (see Section \ref{chem_ions}), this could possibly mean reactions linking these two sets of molecules are badly constrained. This could also indicate that other destruction routes for HCN and HNC are needed to explain the observed column densities.}

{ The bottom panels of Figure \ref{result_nautilus} present the column densities of HCO$^+$, N$_2$H$^+$, and CO. The observed column density of HCO$^+$ (blue area) and N$_2$H$^+$ (red area) are well fitted with model EA1 (both solutions 1 and 2) for all three different initial densities (models 1, 2 and 3). N$_2$H$^+$ is also reproduced with model 4, 5, and 6 (using model EA2) but HCO$^+$ is under-predicted by a factor of $\sim$10 for model 6 and $\lesssim$3 for model 4 and 5. The CO column density is under-estimated for both model EA1 and EA2, by a factor <6 and <3 respectively. Note that the observed CO column density shown in this figure is derived from the C$^{17}$O observations of L1544 derived by \citet{caselli2002-1}.}

{ Table \ref{summary_model} presents a summary of the modelled and observed column densities agreement for all models and all molecules, confronted to the three solutions. For each molecule and each model, all predicted column densities that falls within a factor of 3 in the observed error bars is marked with a cross. From this table, we can conclude that models 1 and 2 associated to solution 1 are the best models to reproduce the observed column densities. These models corresponds to ambient cloud densities of $1\times10^2$\,cm$^{-3}$ and $3\times10^3$\,cm$^{-3}$, and to abundances from model EA1.
One must also note that the model 6 associated to solution 3 might also be a good solution, except for HCN which is over-estimated by a factor of $\sim$30 for this model.}

\begin{table}
	\centering
	\caption{{ Summary of the modelled and observed column densities agreement for all models and all molecules, confronted to the three possible solutions for the age of phase 2: solutions 1 and 2 for models 1, 2, and 3 and solution 3 for model 4, 5, and 6 (see text and Table \ref{init_dens}). A cross indicates that the modelled column density reproduce the observed value within a factor of 3.\label{summary_model}}}
	\begin{tabular}{lccccccccc}
	\hline\hline
			&			\multicolumn{9}{c}{Model number}\\
			&			1&2&3					&	1&2&3					&	4&5&6\\
	\hline
	Molecule 	&			\multicolumn{3}{c}{Solution 1}	&	\multicolumn{3}{c}{Solution 2}	&	\multicolumn{3}{c}{Solution 3}\\
	\hline
	HCNH$^+$		&	X&X&X					&	X&X&X					&	X&X&X\\
	HC$_3$NH$^+$	&	X&X&					&	&&						&	&X&X\\
	HCN				&	X&X&X					&	X&X&					&	&&\\
	HNC				&	X&X&X					&	X&X&X					&	&&X\\
	HC$_3$N			&	X&X&X					&	&X&						&	&X&X\\
	HCO$^+$			&	X&X&X					&	X&X&X					&	&X&X\\
	N$_2$H$^+$		&	X&X&X					&	X&X&X					&	X&X&X\\
	CO				&	X&X&					&	&X&						&	X&X&X\\
	\hline
	Total				&	8&8&6					&	5&7&4					&	3&6&7\\
	\hline
	\end{tabular}
\end{table}

\begin{figure*}
	\centering
	\includegraphics[width=0.49\hsize,clip=true,trim=0 0 0 0]{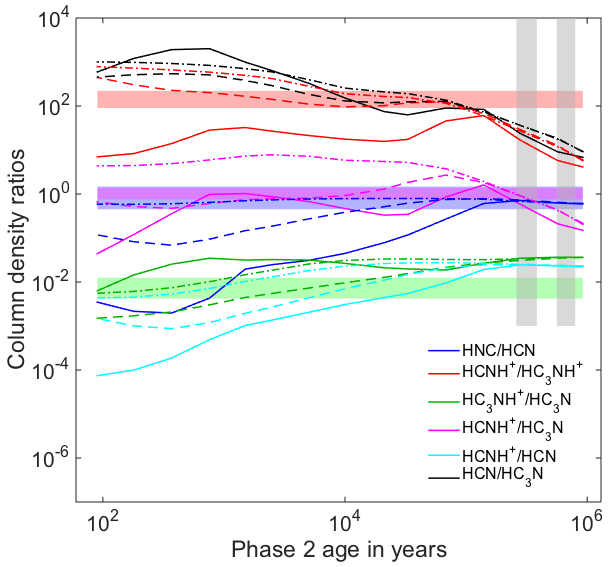}
	\includegraphics[width=0.49\hsize,clip=true,trim=0 0 0 0]{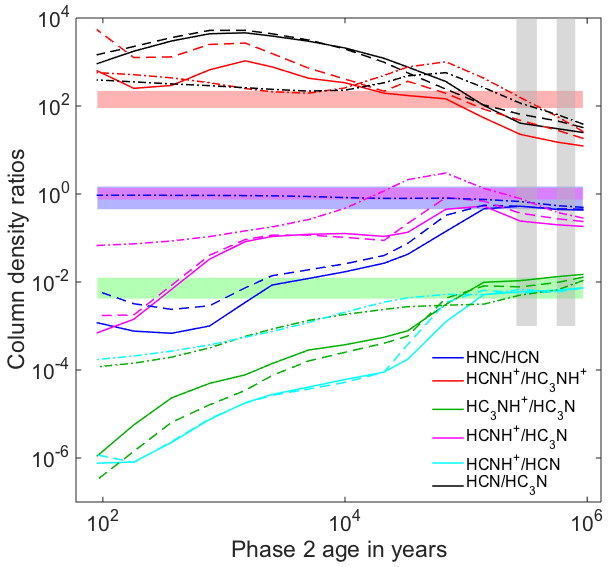}
	\caption{{ Evolution of the column density ratios of HNC/HCN, HCNH$^+$/HC$_3$NH$^+$, HC$_3$NH$^+$/HC$_3$N, HCNH$^+$/HC$_3$N, HCNH$^+$/HCN and HCN/HC$_3$N as a function of the age of the phase 2 for different initial gas densities of the ambient cloud phase: $1\times10^2$\,cm$^{-3}$ (full line), $3\times10^3$\,cm$^{-3}$ (dashed line), and $2\times10^4$\,cm$^{-3}$ (dash-dotted line). The area corresponds to the observed ratio coming from this study and \citet{hily-blant2010} (c.f. text). Grey areas shows the timespan area of confidence of each model based on the observed HCNH$^+$. \textit{Left panels:} Model EA1 of initial atomic abundances. \textit{Right panels:} Model EA2 of initial atomic abundances.}}
	\label{abund_ratios}
\end{figure*}

{ Figure \ref{abund_ratios} presents the column density ratios of HNC/HCN, HCNH$^+$/HC$_3$NH$^+$, HC$_3$NH$^+$/HC$_3$N, HCNH$^+$/HC$_3$N, HCNH$^+$/HCN and HCN/HC$_3$N as a function of time. {The observed HNC/HCN ratio is $0.97\pm0.52$, close to the modelled ratio of $\lesssim$1 for both EA1 and EA2 models.} From our study, we also derived the observed HCNH$^+$/HC$_3$N=$1.1\pm0.3$, HCNH$^+$/HC$_3$NH$^+$=$155\pm65$, and HC$_3$NH$^+$/HC$_3$N=$(8.4\pm4.2)\times10^{-3}$. The predicted ratios with model EA2 gives a good agreement compared to the observed ratios, for an age >10$^5$\,yrs. Model EA1 is worse than model EA2, especially for HCNH$^+$/HC$_3$NH$^+$ and HC$_3$NH$^+$/HC$_3$N. Nonetheless, the HCNH$^+$/HC$_3$N is in good agreement with the observations for both models. This might show that HC$_3$NH$^+$ is slightly better reproduced by model EA2 rather than model EA1.}\\

\subsection{Discussion}

{ To reproduce the observed column densities, several improvement of the chemical network are required, such as:
\begin{enumerate}
	\item A better description of the physical conditions of the source may change the results, by taking into account the collapse of the core between phase 1 and phase 2. Although it may imply other source of error in the modelling.
	\item Formation (or destruction) routes for HCNH$^+$ and HC$_3$NH$^+$ must be added to the chemical network. \citet{agundez2015} also came to this conclusion in their study of the ion NCCNH$^+$ in the cold dark clouds TMC--1 and L483. In their work, using a comparable chemical network (UMIST RATE12, \citealp{mcelroy2013}), the authors reproduced well the abundance of several ions (e.g. NCCNH$^+$, HCO$^+$, NH$_4^+$) but they did not reproduce the observed HCNH$^+$ and HC$_3$NH$^+$ abundances by a factor of $\sim$10, which is similar to the value we find in our study.
	\item A better estimate of reaction rates, e.g. the rate of the dissociative recombination of HC$_3$NH$^+$ with electrons (J.-C. Loison, private communication) and HCNH$^+$ \citep{loison2014-1} where recent rates are not available yet in the KIDA network. 
\end{enumerate}
These points will be developed in a forthcoming study, focused on the complex chemistry of the species involved in this study.}\\
 
\begin{figure*}
	\centering
	\includegraphics[width=0.49\hsize,clip=true,trim=0 0 0 0]{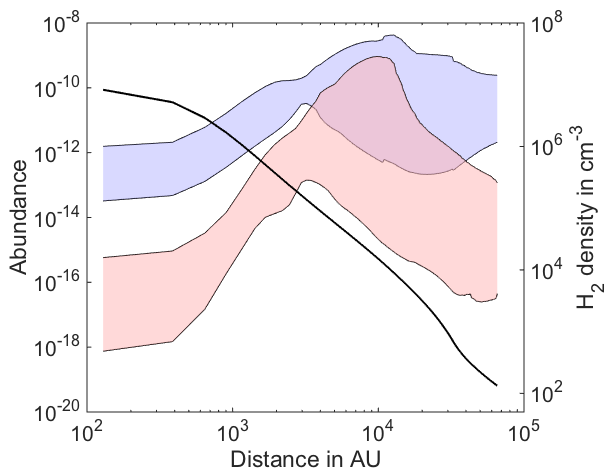}
	\includegraphics[width=0.49\hsize,clip=true,trim=0 0 0 0]{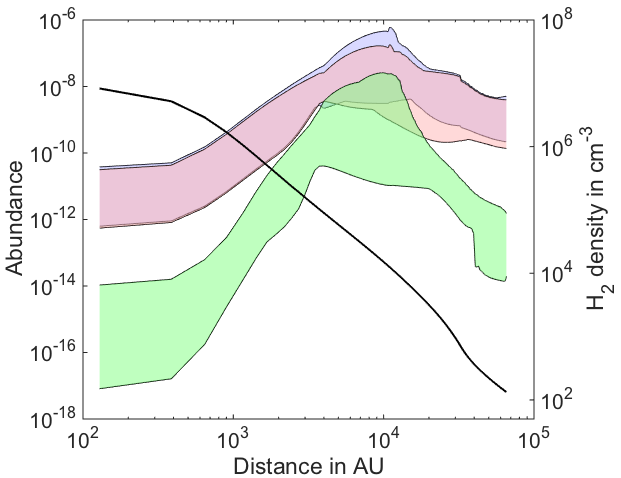}
	\caption{ Evolution of the abundance as a function of the radius. \textit{Left:} HCNH$^+$ (blue) and HC$_3$NH$^+$ (red). \textit{Right:} HCN (blue), HNC (red), and HC$_3$N (green). The boundaries have been obtained combining the different abundance profiles for ages $\gtrsim10^{5}$\,yrs, for both EA1 and EA2 sets. The H$_2$ density profile of L1544 is also shown (black line).}
	\label{abund_radius}
\end{figure*}

Figure \ref{abund_radius} shows the evolution of the abundance of HCNH$^+$ and HC$_3$NH$^+$ as a function of the radius. The area represents the abundance of both species for models with an age $\gtrsim10^{5}$\,yrs. This figure suggests that the emission of these ions is mainly coming from the external part of the core, in a region close to $\sim$10$^{4}$\,au, in an external layer where non-thermal desorption of other species has been previously observed \citep{vastel2014, vastel2016, jimenez-serra2016}. { We have also plotted in Appendix \ref{app1} several figures showing the variation of the abundance, for all models and all eight molecules, as a function of the radius for an age of phase 2 corresponding to solution 1 (3$\times$10$^5$\,yrs).}
 
{ To summarise, we can conclude that for an initial H density of $1\times10^2$ and $3\times10^3$\,cm$^{-3}$, coupled with the model EA1 for elemental abundances, a good agreement can be found for HCNH$^+$, HC$_3$NH$^+$ (and other relevant species) for a phase 2 age larger than $10^5$\,yrs.} This is consistent with the fact that L1544 must be in an advanced pre-stellar stage since it has already begun to collapse \citep{caselli2012, keto2014}.

%%%%%%%%%%%%
\section{Chemistry of HCNH$^+$ and HC$_3$NH$^+$}\label{chem_ions}

The formation of HCNH$^+$ in dense and cold regions such as L1544 is thought to be dominated by both HCN/HNC \citep{schilke1991, hezareh2008, loison2014-1} following:
 \begin{eqnarray}
	\mathrm{H_2 + HCN^+/HNC^+} &\longrightarrow& \mathrm{HCNH^+ + H}\\
	\mathrm{H_3^+ + HCN/HNC} &\longrightarrow& \mathrm{HCNH^+ + H_2}\\
	\mathrm{H_3O^+ + HCN/HNC} &\longrightarrow& \mathrm{HCNH^+ + H_2O}\\
	\mathrm{HCO^+ + HCN/HNC} &\longrightarrow& \mathrm{HCNH^+ + CO}
\end{eqnarray}
Its destruction is dominated by dissociative recombination \citep[e.g.][]{hezareh2008, loison2014-1} following:
 \begin{eqnarray}
	\mathrm{HCNH^+ + \textit{e}^-} &\longrightarrow& \mathrm{HCN + H}\\
	&\longrightarrow& \mathrm{HNC + H}\\
	&\longrightarrow& \mathrm{CN + H + H}
\end{eqnarray}
The two reactions in (2){, involving HCN$^+$ and HNC$^+$, are thought to be the main formation route of HCNH$^+$ in these regions \citep{loison2014-1}. HCN$^+$ and HNC$^+$ are formed by:}
 \begin{eqnarray}
	\mathrm{H^+ + HCN/HNC} &\longrightarrow& \mathrm{HCN^+/HNC^+ + H}\\
	\mathrm{H_2 + CN^+} &\longrightarrow& \mathrm{HCN^+/HNC^+ + H}\\
	\mathrm{H_3^+ + CN} &\longrightarrow& \mathrm{HCN^+/HNC^+ + H_2}\\
	\mathrm{NH_2 + C^+} &\longrightarrow& \mathrm{HCN^+ + H}
\end{eqnarray}
{ Reactions in (9) and (10) are the main formation routes of HCN$^+$ and HNC$^+$ and they are mostly entirely destroyed by reactions (2).}

The chemistry of HC$_3$NH$^+$ is poorly known compared to HCNH$^+$. Several routes to form this species have been proposed \citep{knight1986, kawaguchi1994, osamura1999, takagi1999}:
 \begin{eqnarray}
	\mathrm{C_2H_2^+ + HCN} &\longrightarrow& \mathrm{HC_3NH^+ + H}\\
	\mathrm{C_2H_2^+ + HNC} &\longrightarrow& \mathrm{HC_3NH^+ + H}\\
 	\mathrm{C_2H_2^+ + CN} &\longrightarrow& \mathrm{HC_3N^+ + H}\\
	\mathrm{C_3N^+ + H_2} &\longrightarrow& \mathrm{HC_3N^+ + H}\\
 	\mathrm{HC_3N^+ + H_2} &\longrightarrow& \mathrm{HC_3NH^+ + H}\\
	&\longrightarrow& \mathrm{HCN + C_2H_2^+}
\end{eqnarray}
{ The branching ratio of reactions (17) and (18) are $\sim$\,70\% and $\sim$\,30\% respectively, hence reaction (17) is the main outcome of HC$_3$N$^+$\,$+$\,H$_2$.}
Once HC$_3$NH$^+$ is formed, dissociative recombination will lead to the formation of species HC$_3$N, HNC$_3$, and HCCNC (HCNCC could also be formed using this route but it is the least stable isomer, \citealp{osamura1999}):
 \begin{eqnarray}
	\mathrm{HC_3NH^+ + \textit{e}^-} &\longrightarrow& \mathrm{HC_3N + H}\\
	&\longrightarrow& \mathrm{HNC_3 + H}\\
	&\longrightarrow& \mathrm{HCCNC + H}
\end{eqnarray}

The following reaction might be also involved in the formation of HC$_3$NH$^+$ \citep{federer1986, scott1999}:
 \begin{eqnarray}
	\mathrm{\textit{l}\!-\!C_3H_3^+ + N} &\longrightarrow& \mathrm{HC_3NH^+ + H}.
\end{eqnarray}
This reaction might be important for the formation of HC$_3$NH$^+$ (thus HC$_3$N) but we decided not to include it in the network because it is still subject to discussions and its reaction rate is unknown. A forthcoming paper will study in deeper details the chemistry of both HCNH$^+$ and HC$_3$NH$^+$, including the impact of the latter reaction.

Although the HCN/HNC ratio is predicted to be one via the photodissociation of HCNH$^+$, observations show variations by factors up to three, below or above unity, for the HNC/HCN ratio in dark clouds, and even larger variations in warmer sources \citep{hirota1998}. However, the derivation of the HCN and HNC column densities based on observations of their $^{13}$C isotopologues have been questioned recently by \citet{roueff2015}. It then appears that the deviations of the HNC/HCN ratio from unity may be an observational problem rather than an astrochemical one. The study of the spatial variations of the HCNH$^+$ ion is therefore of utmost importance to elucidate the origin of these variations.

From the Nautilus results, we derived the following reactions as the main production and destruction processes of both species. For HCNH$^+$, as expected from discussed above, the two main production reactions are reactions (2) with both HCN and HNC. In our model, the reaction with HCN (HNC) is responsible for $\sim$60\% ($\sim$30\%) of the production of HCNH$^+$, respectively. The destruction follow the reactions (6), (7), and (8) by a factor of 33\% each, as expected by \citet{semaniak2001} (see also W. D. Geppert's notes in KIDA webiste\footnote{\url{http://kida.obs.u-bordeaux1.fr/datasheet/datasheet_2815_HCNH++e-_V1.pdf}}).
For HC$_3$NH$^+$, it seems to be only produced from reaction (17) (by a factor close to 100\%) and destroyed following reactions (19) and (20) mainly. The respective factor of these reactions is $\sim$56\% and $\sim$32\%. The remaining $\sim$12\% are divided into the production of different isomers such as HCCNC and HCNCC.

Although dissociative recombination with electrons rate constants are well known from experiments \citep{semaniak2001, geppert2004, agundez2013}, formation reaction rates are poorly constrained -- especially for HC$_3$NH$^+$. The reactions rate of (2) and (17), in the KIDA network, comes from estimations. Since these reactions appear to be the main formation pathways of their respective ions, an accurate laboratory measurements of these reactions could determine the precise value of their respective reaction rate, changing the estimated abundance. Moreover, reaction (17) has a branching ratio with reaction (18). A higher ratio for the latter reaction would result in a lower abundance of HC$_3$NH$^+$. Finally, as concluded in the previous section, formation or destruction routes could also be added to this network in order to reproduce the observed column densities.
A lack of laboratory experiments and observational data does not allow, to date, to disentangle which reactions dominate the formation of both HCNH$^+$ and HC$_3$NH$^+$. Reaction rates are not well constrained and it may lead to numerous error in the abundance determination.

%%%%%%%%%%%%%%%%%%%%%%%%%%%%%%%%%
\section{Conclusions}

{ In this paper, we presented the detection of one HFS transition of HCNH$^+$ and three HFS transition of HC$_3$NH$^+$ in the pre-stellar core L1544. An LTE analysis has been performed and we derive N(HCNH$^+$)\,=\,(2.0$\pm$0.2)$\times$10$^{13}$\,cm$^{-2}$ and N(HC$_3$NH$^+$)\,=\,(1.5$\pm$0.5)$\times$10$^{11}$\,cm$^{-2}$. 
We also report the detection of five transitions of HC$_3$N, three transitions of H$^{13}$CN and one transition of HN$^{13}$C. A non-LTE analysis gives a column density N(HC$_3$N) = (2.0$\pm$0.4)$\times$10$^{13}$\,cm$^{-2}$, N(HCN) = (3.6$\pm$0.9)$\times10^{14}$\,cm$^{-2}$, and N(HNC) = (3.0$\pm$1.0)$\times$10$^{14}$\,cm$^{-2}$ with a kinetic temperature of $11\pm3$\,K and a density of (6$\pm$4)$\times$10$^4$\,cm$^{-3}$.
Using a gas-grain chemical code, we calculated the predicted abundance in the pre-stellar core using a two-step model by varying the initial gas densities of the initial phase (within the range found in previously published works). The resulting abundances have been compared to the observed column densities and models with an age $>10^5$\,yrs gives a satisfactory agreement to the observations, in agreement with recent studies of this source.
We have also discussed the possible predominant chemical reactions leading to the formation and destruction of both ionic species. Further investigations of the chemistry of nitrogen species (with laboratory experiments) is required to better reproduce the observations, especially for HC$_3$NH$^+$.}

\section*{Acknowledgements}
The authors want to acknowledge J.-C. Loison and V. Wakelam for fruitful conversations about the chemical network { and modelling} of the species. This study is based on observations carried out as part of the Large Program ASAI (project number 012--12) with the IRAM 30m telescope. IRAM is supported by INSU/CNRS (France), MPG (Germany) and IGN (Spain). This work was supported by the CNRS program ``Physique et Chimie du Milieu Interstellaire'' (PCMI) and by a grant from LabeX Osug\@2020 (Investissements d'avenir - ANR10LABX56). D. Q. acknowledges the financial support received from the STFC through an Ernest Rutherford Grant (proposal number ST/M004139/2).

\bibliographystyle{mnras}
\bibliography{All_ref} % if your bibtex file is called example.bib

%%%%%%%%%%%%%%%%%%%%%%%%%%%%%%%%%%%%%%%%%%%%%%%%%%

%%%%%%%%%%%%%%%%% APPENDICES %%%%%%%%%%%%%%%%%%%%%

\appendix

\section{Abundance profiles}\label{app1}

{ We present in this appendix the abundance profiles for all six models as a function of the radius for an age of phase 2 of 3$\times$10$^5$\,yrs, for all eight molecules.
This age corresponds to solution 1, the best solution as determined by Table \ref{summary_model}. For all eight molecules, models 1 and 2 reproduce the observed column densities within a factor of 3.

One can note that the modelled peak abundance for HCN, HNC, and HC$_3$N is located at a radial distance of $\sim$\,[4--5]$\times$10$^3$\,au. At this distance, from Fig. \ref{struct_l1544}, the kinetic temperature is 10\,K with a H$_2$ density of $\sim$\,$1\times10^5$\,cm$^{-3}$. Both value are consistent with the non-LTE modelling performed in Sect. \ref{results} for these species.}

\begin{figure*}
	\centering
	\includegraphics[width=0.45\hsize,clip=true,trim=0 0 0 0]{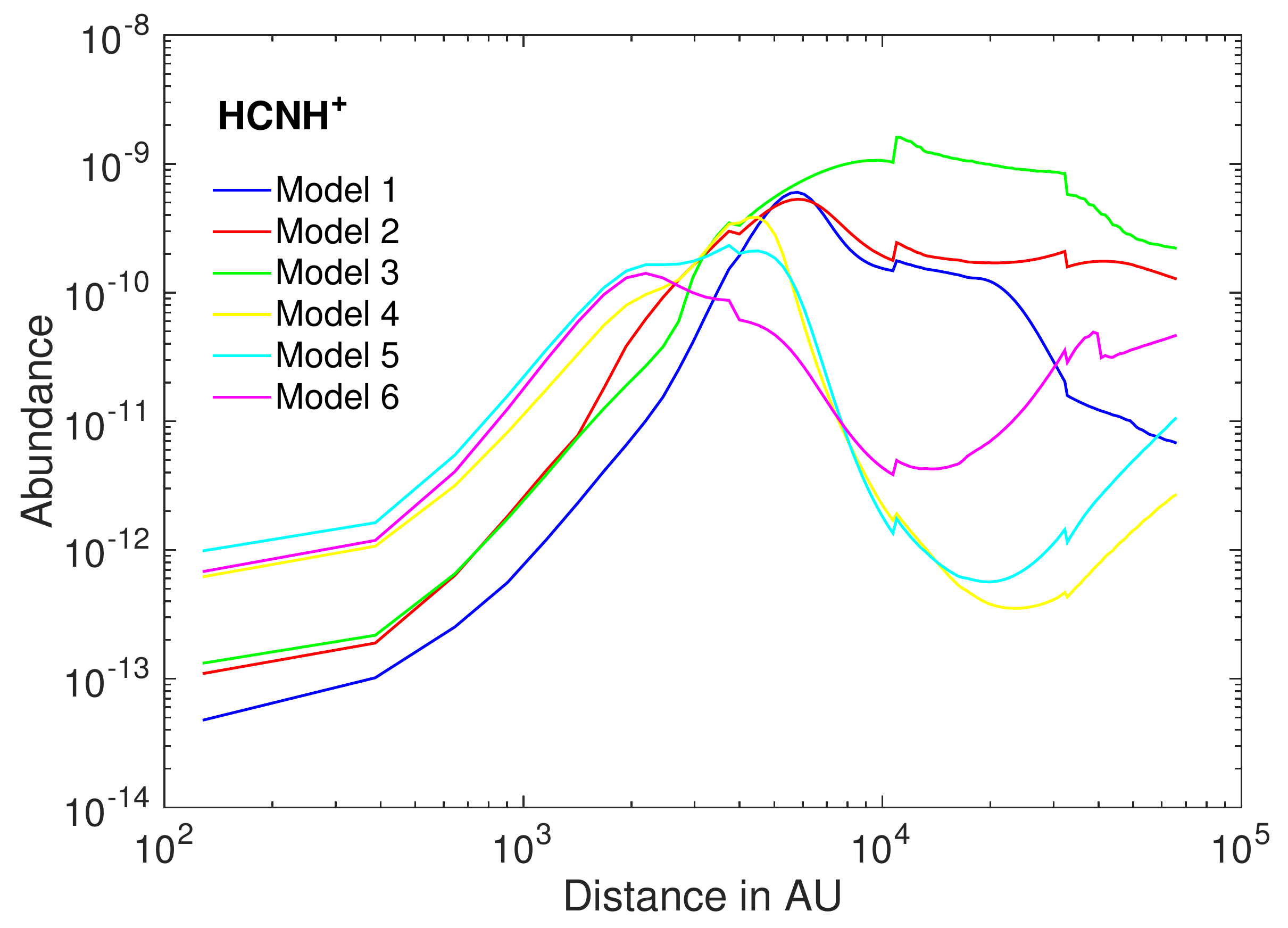}
	\includegraphics[width=0.45\hsize,clip=true,trim=0 0 0 0]{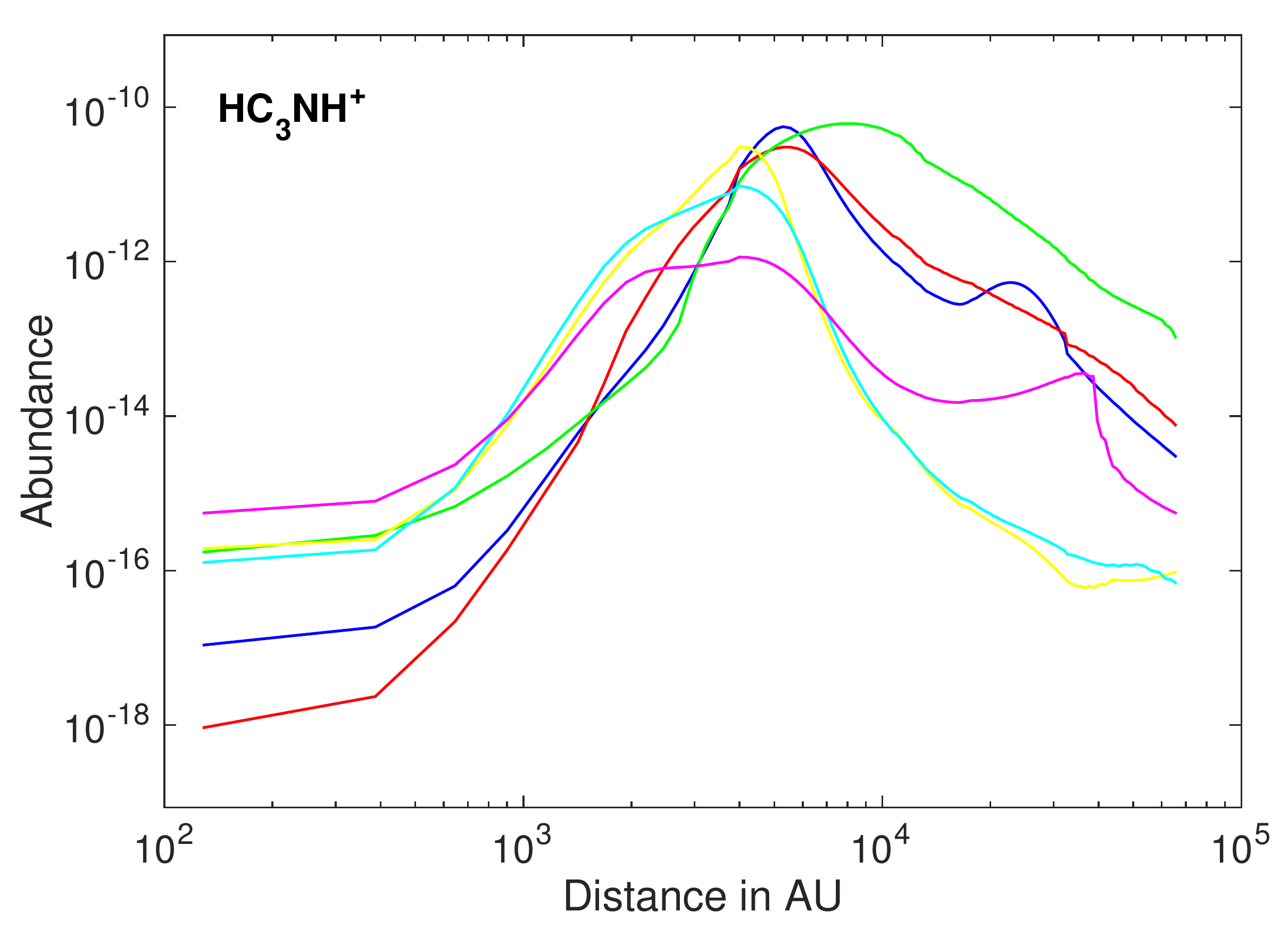}\\
	\includegraphics[width=0.45\hsize,clip=true,trim=0 0 0 0]{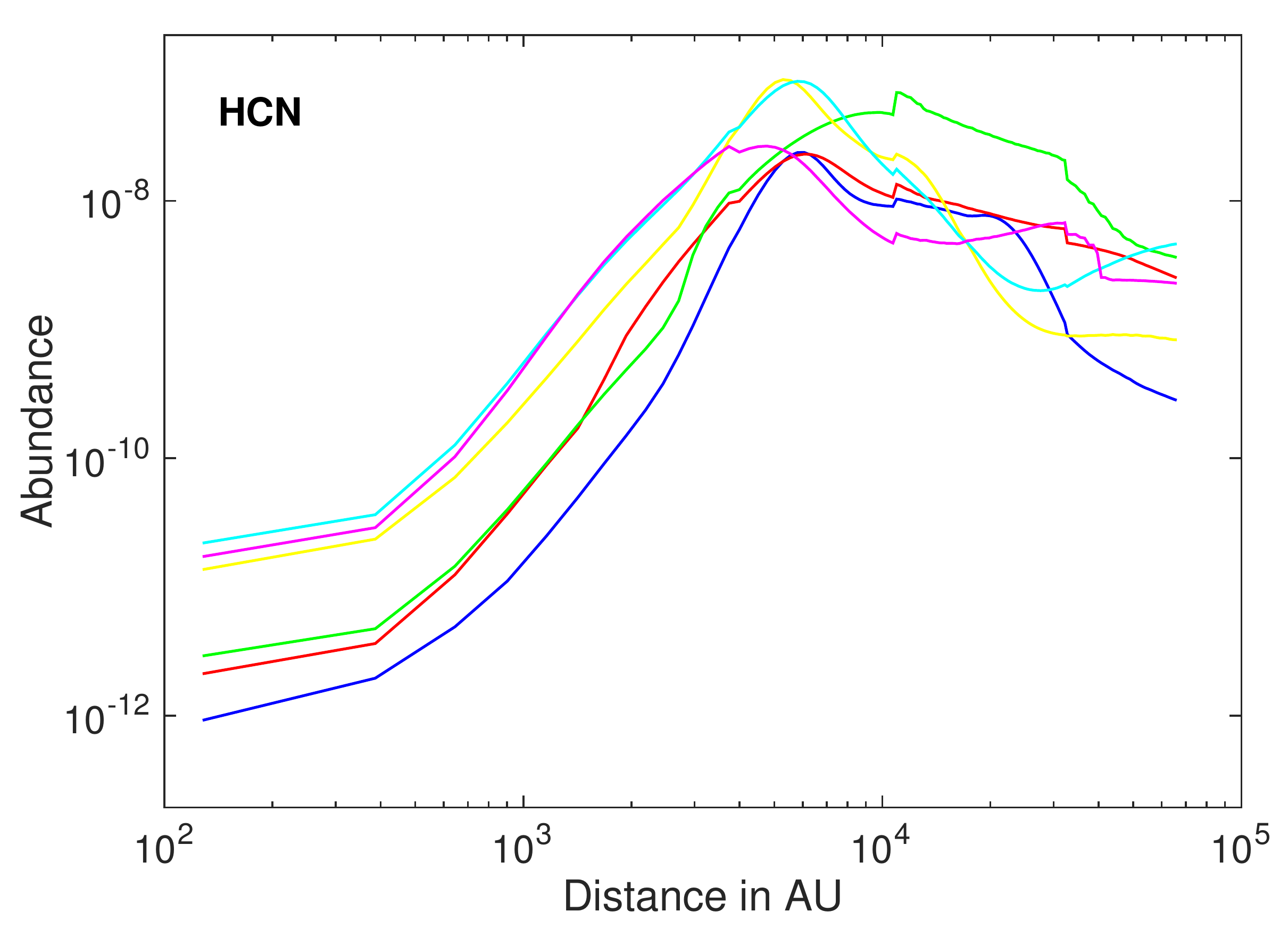}
	\includegraphics[width=0.45\hsize,clip=true,trim=0 0 0 0]{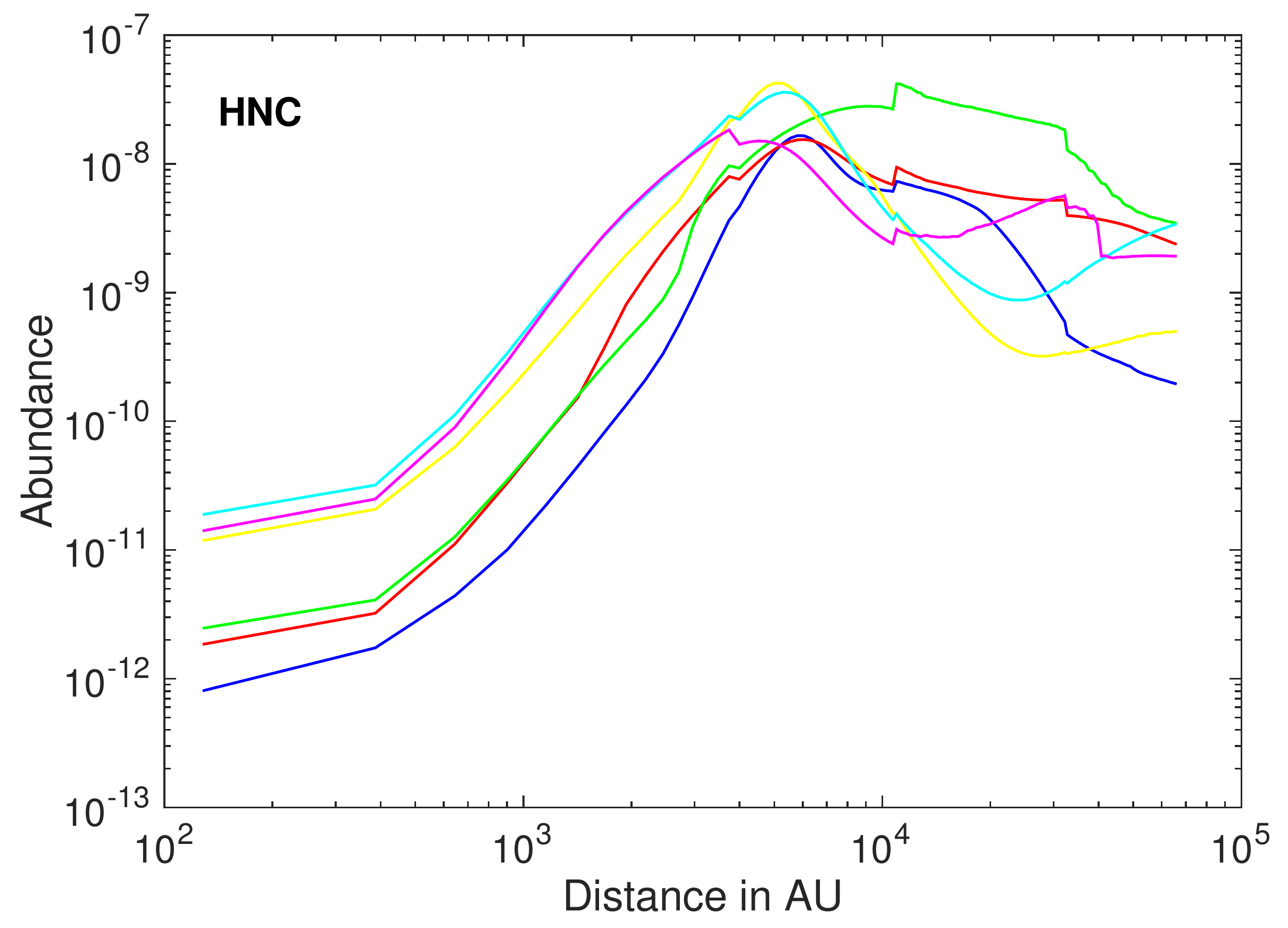}\\
	\includegraphics[width=0.45\hsize,clip=true,trim=0 0 0 0]{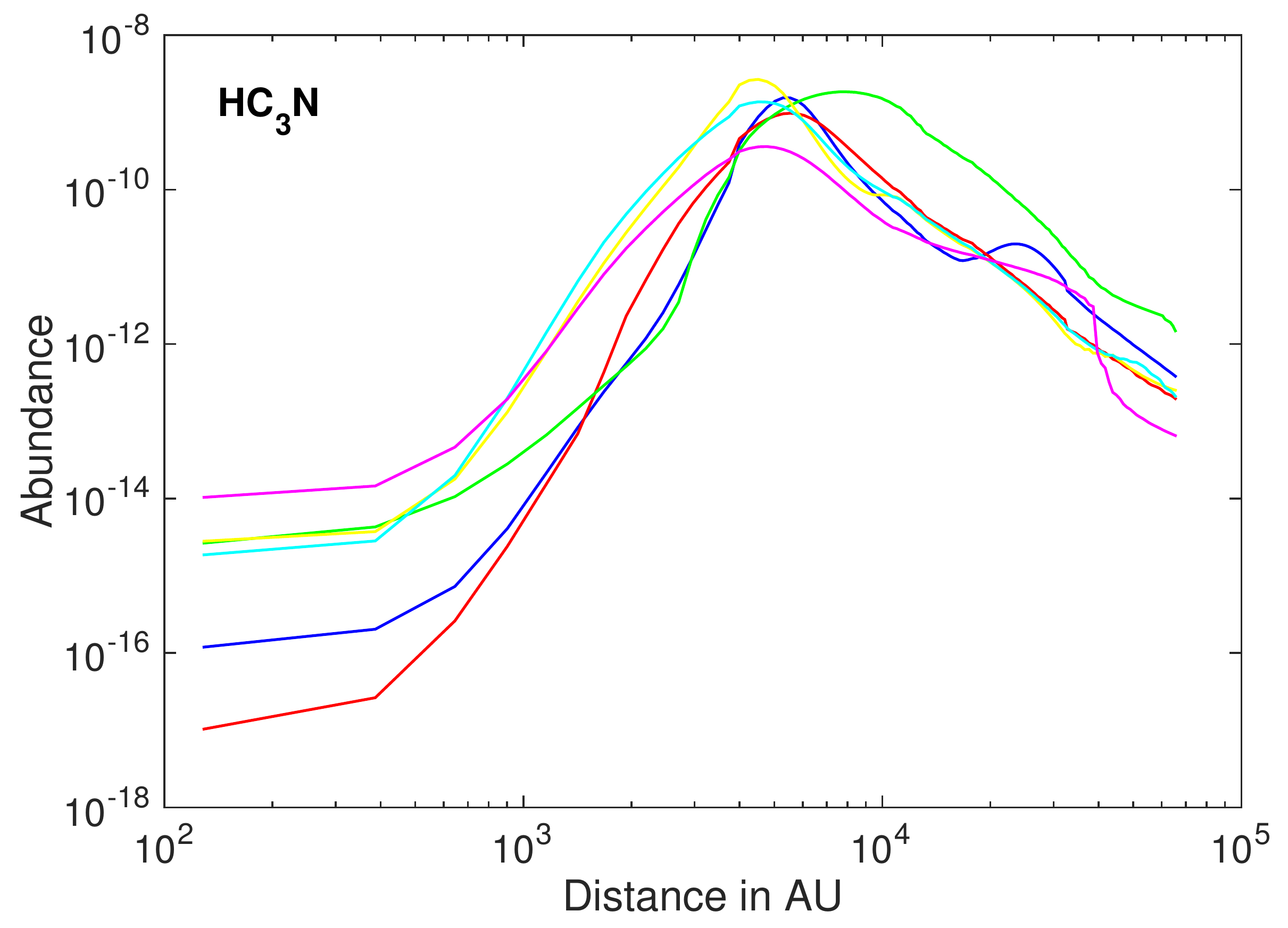}
	\includegraphics[width=0.45\hsize,clip=true,trim=0 0 0 0]{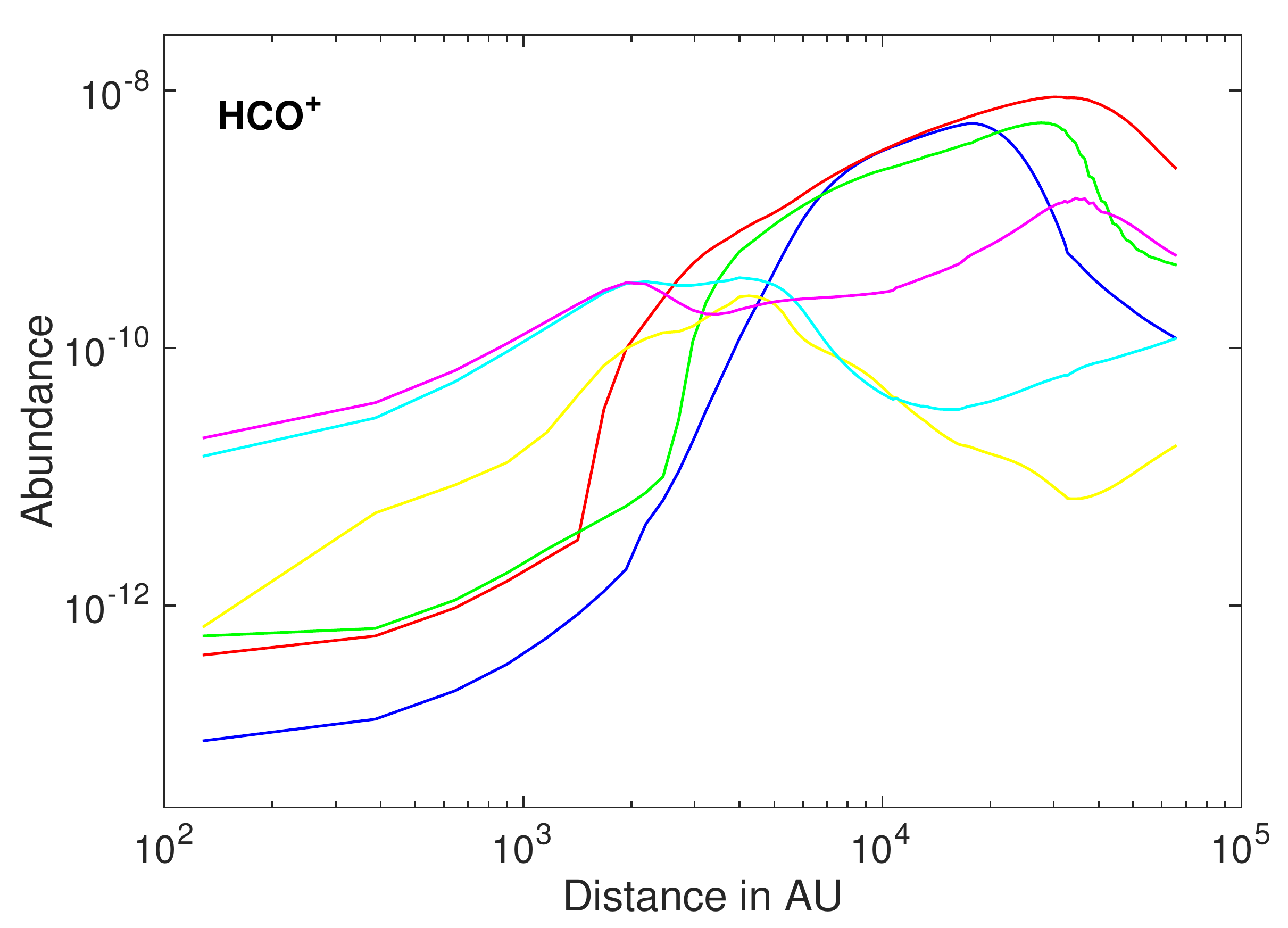}\\
	\includegraphics[width=0.45\hsize,clip=true,trim=0 0 0 0]{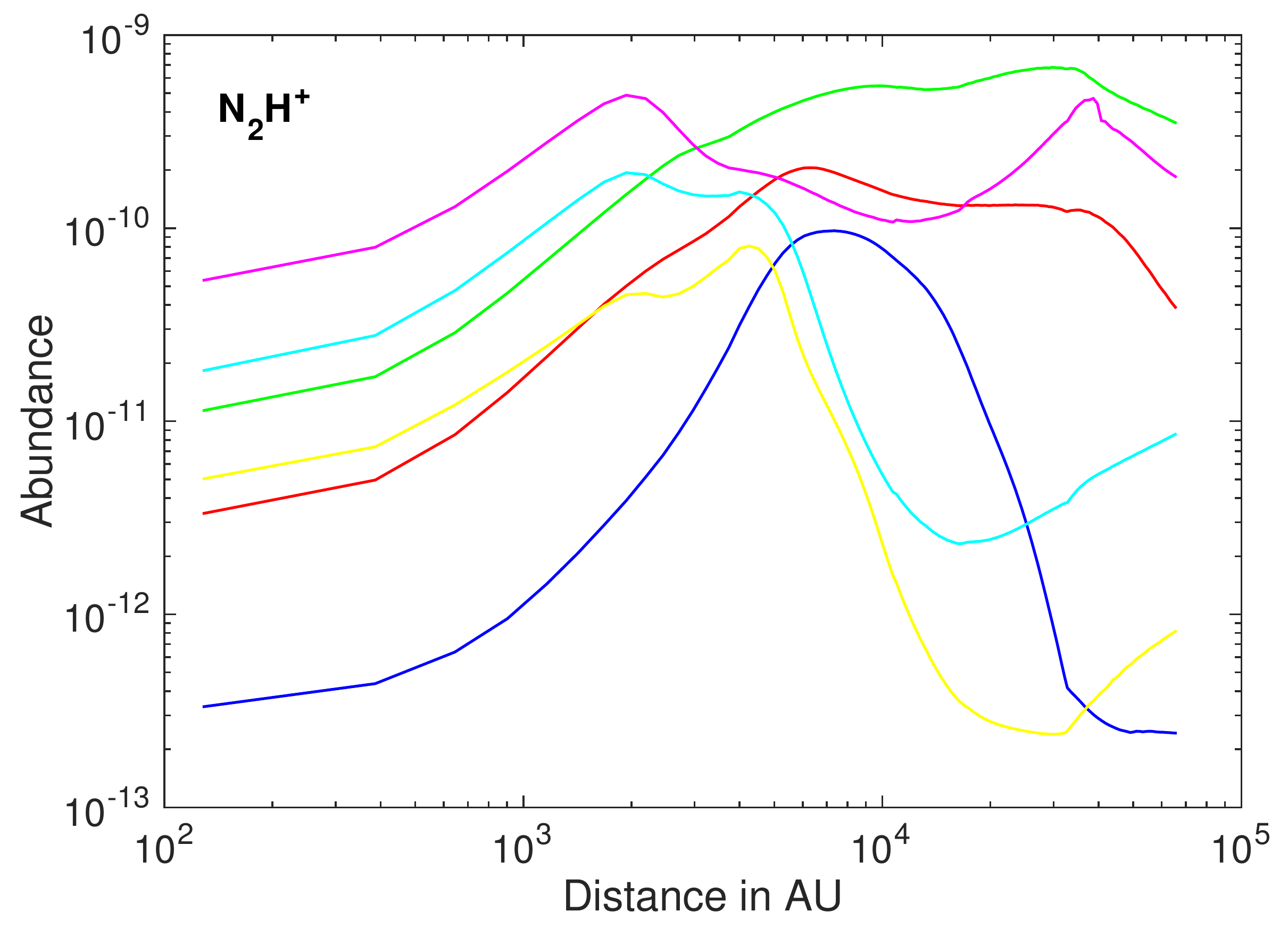}
	\includegraphics[width=0.45\hsize,clip=true,trim=0 0 0 0]{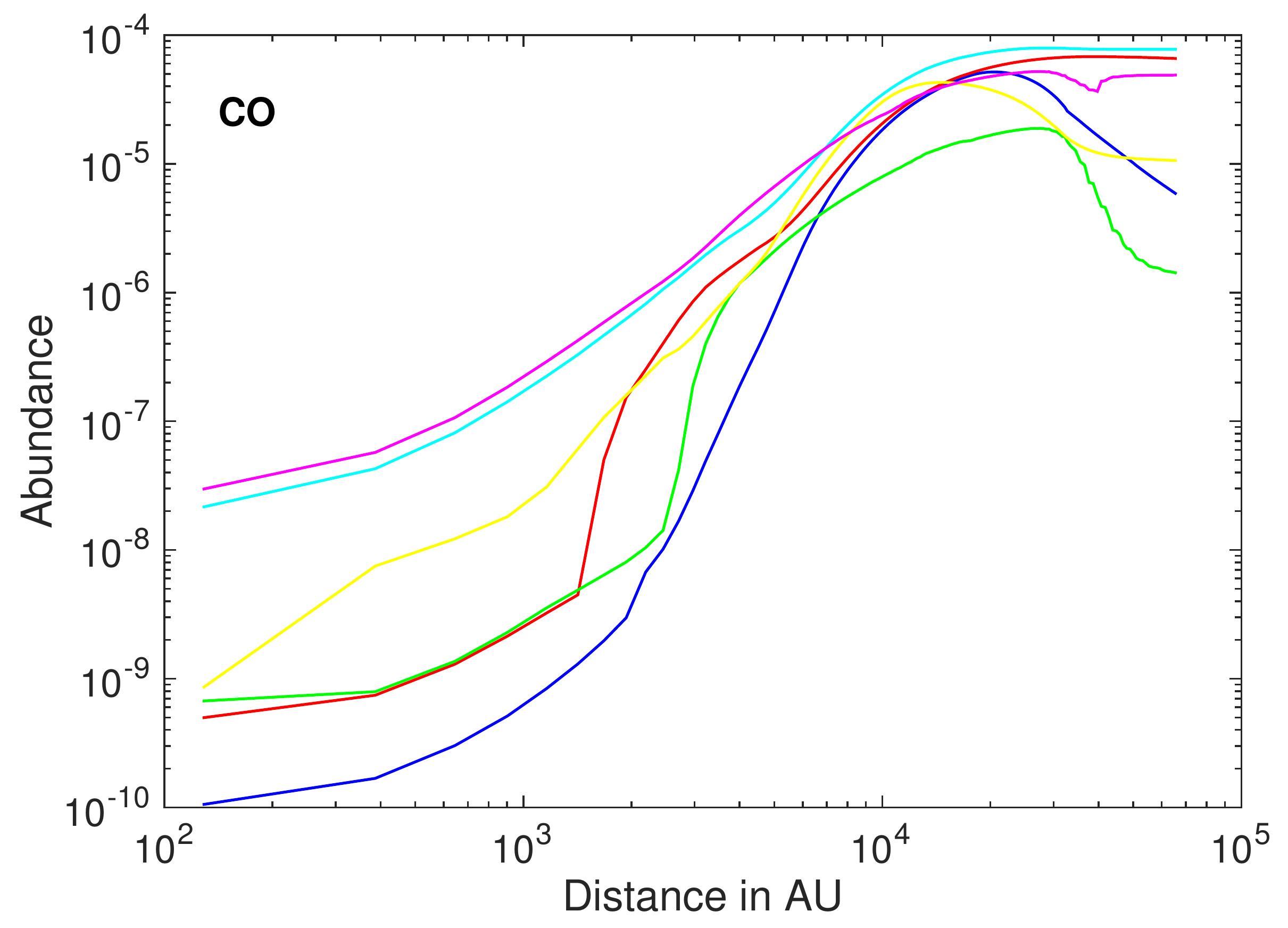}\\
	\caption{ Abundance profiles as a function of the radius given for all models for an age of phase 2 of 3$\times$10$^5$\,yrs.}
	\label{abund_model_2}
\end{figure*}

%%%%%%%%%%%%%%%%%%%%%%%%%%%%%%%%%%%%%%%%%%%%%%%%%%

% Don't change these lines
\bsp	% typesetting comment
\label{lastpage}
\end{document}